\documentclass[12pt]{article}
\pdfoutput=1
\usepackage{makeidx}
\makeindex
\usepackage[a4paper]{geometry}
\usepackage{jheppub,amsmath,amssymb,amsfonts,amsxtra,mathrsfs,graphics,graphicx,amsthm,epsfig,ytableau,bm,longtable,float,color,tikz,mathtools,xfrac,footnote,rotating,lscape}
\usepackage{dsfont}
\usepackage{textgreek}
\usetikzlibrary{matrix,decorations}
\usetikzlibrary{decorations.pathmorphing}
\usetikzlibrary{decorations.markings}
\usetikzlibrary{arrows, decorations.markings, calc, fadings, decorations.pathreplacing, patterns, decorations.pathmorphing, positioning}
\usepackage{tikz-cd}
\usepackage{tikz-3dplot}

\usepackage{stackengine}
\stackMath
\usepackage{mathtools}
\usepackage{esint}
\usepackage[inline]{enumitem}
\usepackage{fixmath} 
\usepackage{scalerel}
\newlength\bshft
\def\fakebold#1{\ThisStyle{\ooalign{$\SavedStyle#1$\cr%
  \kern-\bshft$\SavedStyle#1$\cr%
  \kern\bshft$\SavedStyle#1$}}}

\usetikzlibrary{positioning,shapes}
\usetikzlibrary{chains}
\usetikzlibrary{arrows,fit,decorations.pathreplacing}
\tikzstyle{every picture}+=[remember picture]
\tikzstyle{na} = [baseline=-.5ex]

\usepackage{empheq}
\usepackage{multirow}
\usepackage{booktabs}
\usepackage[american]{babel}

\usepackage[latin1]{inputenc}

\usepackage[inline]{enumitem}

\makeatletter
\newcommand{\vast}{\bBigg@{1}}
\newcommand{\Vast}{\bBigg@{5}}
\makeatother

\usepackage{hyperref}

\usepackage{subfigure}

\numberwithin{equation}{section}

\newcommand{\tv}{\texttt{v}}

\usepackage[bbgreekl]{mathbbol}
\DeclareMathSymbol\bbDelta  \mathord{bbold}{"01}

\newcommand{\eg}{\textit{e.g.}}

\newcommand{\ii}{\mathrm{i}}
\newcommand{\?}{\;\!}

\numberwithin{equation}{section}

\newcommand{\be}{\begin{equation}} \newcommand{\ee}{\end{equation}}
\newcommand{\bea}{\begin{equation} \begin{aligned}} \newcommand{\eea}{\end{aligned} \end{equation}}

\def\U{\mathrm{U}}

\newcommand{\rd}{\mathrm{d}}

\newcommand{\wt}{\widetilde}

\newcommand{\pd}{\partial}

\DeclareMathOperator{\Tr}{Tr}

\DeclareMathOperator{\sign}{sign}

\DeclareMathOperator{\re}{\mathbb{R}e}

\DeclareMathOperator{\Li}{Li}

\newcommand{\cC}{\mathcal{C}}

\newcommand{\cG}{\mathcal{G}}

\newcommand{\cI}{\mathcal{I}}

\newcommand{\cN}{\mathcal{N}}
\newcommand{\cO}{\mathcal{O}}

\newcommand{\cV}{\mathcal{V}}
\newcommand{\cW}{\mathcal{W}}

\newcommand{\bB}{\mathbb{B}}

\newcommand{\bH}{\mathbb{H}}

\newcommand{\bZ}{\mathbb{Z}}

\newcommand{\fg}{\mathfrak{g}}

\newcommand{\fh}{\mathfrak{h}}

\newcommand{\fm}{\mathfrak{m}}

\newcommand{\fn}{\mathfrak{n}}

\newcommand{\fp}{\mathfrak{p}}

\newcommand{\fR}{\mathfrak{R}}
\newcommand{\fs}{\mathfrak{s}}
\newcommand{\ft}{\mathfrak{t}}

\usepackage[Symbol]{upgreek}
\usepackage{bm}

\usepackage{calligra}
\DeclareMathAlphabet{\mathcalligra}{T1}{calligra}{m}{n}

\theoremstyle{plain}

  \theoremstyle{definition}

\providecommand{\examplename}{Example}
\providecommand{\theoremname}{Theorem}

\makeatletter
\g@addto@macro\bfseries{\boldmath}
\makeatother

\usepackage{changes}
\definechangesauthor[name={Per cusse}, color=orange]{per}

\title{The large $N$ limit of topologically twisted indices: a direct approach}
\author[a]{Seyed Morteza Hosseini}
\author[b,c]{and Alberto Zaffaroni}
\affiliation[a]{Department of Physics, Imperial College London, London, SW7 2AZ, UK}
\affiliation[c]{INFN, sezione di Milano-Bicocca, I-20126 Milano, Italy}
\affiliation[d]{Dipartimento di Fisica, Universit\`a di Milano-Bicocca, I-20126 Milano, Italy}
\emailAdd{s.hosseini@imperial.ac.uk}
\emailAdd{alberto.zaffaroni@mib.infn.it}

\abstract{We study the large $N$ limit of the refined topologically twisted index of three-dimensional $\mathcal{N} = 2$ gauge theories with a holographic dual, which is relevant for counting the number of microstates of rotating, magnetically charged and twisted AdS$_4$ black holes. As a difference with previous computations, we perform a direct saddle point analysis of the relevant matrix model. Among other things we explicitly identify the distribution of gauge magnetic fluxes that contributes at large $N$. The index exhibits a large $N$ factorized form in agreement with the supergravity expectations based on gravitational blocks and correctly reproduces the entropy of magnetically charged rotating AdS$_4 \times S^7$ black holes.}

\begin{document}

\setcounter{tocdepth}{2}
\maketitle

%
%

\date{Dated: \today}




\section{Introduction}

The topologically twisted index  is the supersymmetric partition function  on $\Sigma_\fg \times S^1$ of  three-dimensional $\cN = 2$ gauge theories  with
 a topological twist along the Riemann surface $\Sigma_\fg$ \cite{Benini:2015noa}. It can be defined as the equivariant Witten index 
 \bea Z_{\Sigma_\fg \times S^1}(y_I, \fs_I) = \Tr_{\Sigma_\fg} (-1)^F e^{-\beta \{Q,Q^\dagger\}} \prod_I y_I^{J_I} \, ,\eea
 of the theory compactified on $\Sigma_\fg$, where $\fs_I$ are magnetic fluxes on the Riemann surface specifying the twist and $y_I$ are complexified fugacities for the flavor symmetries $J_I$. 
 For theories with an AdS$_4$ holographic dual, the index is supposed to count the number of microstates
 of magnetically charged and topologically twisted asymptotically AdS$_4$ black holes. The entropy of the most general family of static  twisted black holes in AdS$_4\times S^7$ has been  reproduced in \cite{Benini:2015eyy} by studying the large $N$ limit of the topologically twisted index of the dual ABJM theory \cite{Aharony:2008ug}.\footnote{This analysis has been extended to many other twisted black objects in AdS$_4$ and in higher dimensions. For a (partial) review, see \cite{Zaffaroni:2019dhb}.} 

In this paper we study the large $N$ limit of the  \emph{refined} topologically twisted index  of three-dimensional $\cN = 2$ gauge theories with a holographic dual. This index can be defined only when $\fg=0$ and it contains a further refinement with respect to rotational $\U(1)$ symmetry of the two sphere. It corresponds to a background $S^1 \times S^2_\epsilon$, where $\epsilon$ is the equivariant parameter for the rotation and it specifies the $S^1$ fibration over $S^2$. Holographically, it is supposed to count the number of microstates  of the magnetically charged rotating AdS$_4 \times S^7$ black holes found in \cite{Hristov:2018spe} and their generalizations.

 The original method used in \cite{Benini:2015eyy}, based on a decomposition of the partition function in a sum over Bethe vacua, cannot be applied to the refined index. In this paper we will take a different approach by  evaluating directly the large $N$ limit of the index matrix model. Localization reduces the index to a finite-dimensional integral over  gauge holonomy variables, summed over topological sectors specified by gauge magnetic fluxes along $S^2$. We will assume that in the large $N$ limit the fluxes can be treated as continuous variables and  we will take a saddle point approximation with respect  to both  holonomies and  fluxes.  The method has been already efficiently used to study five-dimensional indices \cite{Hosseini:2018uzp,Jain:2021sdp,Hosseini:2021mnn}.  We will of course reproduce the results in \cite{Benini:2015eyy} and their generalizations \cite{Hosseini:2016tor,Hosseini:2016ume} for $\epsilon=0$ and we will provide the general result
 for generic $\epsilon$. One of the advantages of the direct method is that it provides the distribution of magnetic fluxes that dominates the index at large $N$. This can be useful
for understanding the quantum mechanics obtained by a dimensional reduction on the sphere \cite{Benini:2022}.

A different method, valid in the Cardy limit $(\epsilon \to 0)$,  was proposed in  \cite{Choi:2019dfu} and successfully applied to the $\cN=8$ theory coupled to a fundamental hypermultiplet, which is supposed to flow to ABJM in the infrared.
We will explicit discuss the relation between the two methods.

One of the results of our analysis is that, for the class of theories we will consider, the large $N$ refined index, to all perturbative orders in $\epsilon$, can be written in a factorized form 
\be\label{ADHMinndexfac0000}
 \log Z = \ii \sum_{\sigma = 1}^2 \frac{\mathring \cW ( \Delta^{(\sigma)} )}{\epsilon^{(\sigma)}} \, ,
\ee
where $y_I=e^{\ii \Delta_I}$ and $\mathring{\cW} ( \Delta )$ is the on-shell twisted effective potential defined in section \ref{sec:3} and proportional
to the $S^3$ free-energy of the $\cN=2$ gauge theory at large $N$. The explicit form of the gluing in given in \eqref{A-twisted:var}. The result
\eqref{ADHMinndexfac0000} is in complete agreement with the general holographic expectations based on  gravitational blocks \cite{Hosseini:2019iad}
and correctly reproduce the entropy of the rotating AdS$_4 \times S^7$ black holes found in \cite{Hristov:2018spe}. Similar large $N$ factorizations hold in higher dimensions too \cite{Hosseini:2021mnn}. 
In this paper we work to all orders but perturbatively in $\epsilon$. A more refined analysis would be needed 
to understand if there are exponentially suppressed  corrections in $\epsilon$ to the result \eqref{ADHMinndexfac0000}. 

The paper is organized as follows. In section \ref{sec:RTTI} we review the localization formula for the refined topologically twisted index. In \ref{sec:3} we give a general overview of the available methods for computing the topologically twisted inded both for $\epsilon=0$ and $\epsilon \ne 0$. In particular we point out the differences between the Bethe route used in \cite{Benini:2015eyy} and the other direct methods. A particular interesting technical point is the difference in the treatment of the so-called tails regions of the eigenvalues distribution. The interested reader will find details in section \ref{sec:3} and appendix \ref{app:N3/2}. In section \ref{sec:4} we present the general rules for computing the index  for a class of quivers with $N^{3/2}$ scaling of the free-energy and holographically dual to AdS$_4\times M_7$ M-theory backgrounds, where $M_7$ is a seven-dimensional Sasaki-Einstein manifold. We will explicitly present two examples, the ADHM quiver and the ABJM theory. In section \ref{sec:5} we present the general rules for computing the index  for a class of quivers with $N^{5/3}$ scaling of the free-energy and holographically dual to AdS$_4$ backgrounds in massive type IIA. A series of appendices contains the technical derivations and explicit formulae for ADHM and ABJM solutions.

\section{Refined twisted index for 3d $\cN = 2$ field theories}
\label{sec:RTTI}

Consider a three-dimensional $\cN = 2$ gauge theory with gauge group $\cG$,
$I$ chiral multiplets in a representation $\oplus \fR_I$ of the gauge group, and Chern-Simons couplings $k_a$ for the various factors $\cG^{(a)}$ of the group $\cG=\prod_a \cG^{(a)}$.
The \emph{refined} topologically twisted index reads \cite{Benini:2015noa}
\be
 \label{RTTI:main}
 Z_{S^1 \times S^2_\epsilon} ( \Delta , \fs, \Delta_m , \fs_m | \epsilon ) = \frac{1}{|\mathfrak{W}|} \sum_{ \fm \in \Gamma_\fh}
 \oint_{\text{JK}} \prod_{i = 1}^{\text{rk}(\cG)}\frac{\rd x_i}{2 \pi \ii x_i}
 Z_{\text{cl}} ( u_i, \fm_i  ; \Delta_\fm , \fs_m )
 Z_{1\text{-loop}} ( u_i, \fm_i ; \Delta, \fs | \epsilon ) \, ,
\ee
where
\bea
 \label{Z:1-lopp}
 Z_{1\text{-loop}} ( u_i, \fm_i ; \Delta, \fs | \epsilon ) & =
 q^{- \frac12 \sum_{\alpha > 0} | \alpha ( \fm ) |} \prod_{\alpha \in \cG} \left( 1 - x^\alpha q^{| \alpha ( \fm ) | / 2} \right) \\
 & \times \prod_{I} \prod_{\rho_I \in \fR_I} \prod_{\ell = - \frac{| B_I | - 1}{2}}^{\frac{| B_I | - 1}{2}}
 \left( \frac{x^{\rho / 2} y^{\nu / 2} q^{\ell / 2} }{1 - x^{\rho} y^{\nu} q^{\ell}} \right)^{\sign (B_I)} \, .
\eea
Here, $B_I \equiv \rho_I ( \fm ) - \nu ( \fs ) + 1$, $\alpha$ denotes the roots of the gauge group, $\rho$, $\nu$ are the weights of the chiral multiplet under the gauge and flavor symmetry groups, respectively, and $|\mathfrak{W}|$ is the order of the Weyl group of $\cG$.
In this formula, $( \fm, \fs )$ are the gauge and flavor magnetic fluxes on $S^2_\epsilon$, respectively;
$x = e^{\ii u}$, $y = e^{\ii \Delta}$ are the gauge and flavor fugacities  and $q = e^{\ii \epsilon}$ is the fugacity for the angular momentum refinement. 

The classical contributions come from the Chern-Simons terms and the topological symmetries.  A factor $\cG^{(a)}$ of the gauge group $\cG$  contributes
\bea  Z_{\text{cl}}^{\text{CS}} ( u_i, \fm_i )  = \prod_{i = 1}^{\text{rk} (\cG^{(a)})} x_i^{\ii k_a \fm_i}   \, , \eea
where $k_a$ is the associated Chern-Simons coupling. A topological symmetry  associated with a  $\U(1)$ factor  contributes 
\bea
 Z_{\text{cl}}^{\text{top}} ( u, \fm ; \Delta_\fm , \fs_m ) = x^{\fs_m} \zeta^\fm \, ,
\eea
where $\fs_m$ is the topological magnetic flux and $\zeta = e^{\ii\Delta_m}$ is the topological fugacity.
Observe that the classical contribution $Z_{\text{cl}} ( u_i, \fm_i )$  does \emph{not} depend on the refinement parameter $\epsilon$.

\section{Matrix model large $N$ limit}\label{sec:3}

We can take various approaches to evaluate the large $N$ limit of the refined index \eqref{RTTI:main}.
Each has its own virtues that we are going to briefly describe in the following. 

\subsection{The Bethe route}\label{Betheroute}

The large $N$ limit of the \emph{unrefined} index,  \eqref{RTTI:main} for $\epsilon=0$, was first considered in
\cite{Benini:2015eyy} for the ABJM theory, and later generalized in \cite{Hosseini:2016tor,Hosseini:2016ume} to more general Chern-Simons quivers with gauge group $\cG = \prod_{a = 1}^{|\cG|} \U(N)_a$ and
matter transforming in the bi-fundamental, adjoint and (anti-)fundamental representations with a holographic dual.
Further generalizations to other type of quivers relevant for holography can be found in \cite{Jain:2019euv,Jain:2019lqb,Amariti:2019pky,Coccia:2020cku,Coccia:2020wtk}.

The Bethe approach is based on the fact that the unrefined index can be written as a sum of contributions 
\bea
 \label{indexBE}
 Z_{S^1 \times S^2} ( \Delta , \fs | 0 ) =\sum_{u = u^*} e^{-\Omega (u ; \Delta , \fs)} \left (\det_{ij} \partial^2_{u_i u_j} \wt \cW(u ; \Delta) \right )^{-1}\, ,
\eea
where $\wt \cW(u ; \Delta)$ and $\Omega(u ; \Delta , \fs)$ are the effective \emph{twisted superpotential} and the \emph{effective dilaton}  obtained by reducing the three-dimensional theory on $S^1$, 
whose explicit expressions can be found in \cite{Nekrasov:2014xaa,Closset:2017zgf}. The sum is over the Bethe vacua, the critical points of the effective twisted superpotential
\be
 \label{BE}
 \exp \left( \ii \frac{\partial \wt \cW ( u ; \Delta )}{\partial u_i} \right) \bigg|_{u = u^*} =1 \, .
\ee
This form of the index follows from \eqref{RTTI:main} by observing that the sum over gauge magnetic fluxes $\fm_i$ can be explicitly performed when $\epsilon=0$.
This leads to a set of poles in the integrand at the solutions to \eqref{BE} and the residue theorem then gives \eqref{indexBE} \cite{Benini:2015eyy}.
 
The main idea behind the Bethe approach is that one contribution will dominate \eqref{indexBE}  in the large $N$ limit and this can be found by taking the large $N$ limit of \eqref{BE}
\be
 \label{BE2}
 \frac{\partial \wt \cW ( u ; \Delta )}{\partial u_i}   = 2 \pi n_i \, , \qquad \qquad n_i \in \bZ  \, ,
\ee
where the integer $n_i$ are carefully chosen to guarantee the existence of the limit.
For the Chern-Simons gauge theories of interest, as shown in \cite{Benini:2015eyy,Hosseini:2016tor},
the gauge holonomies $u_i$ at the saddle point are distributed along a curve in the complex plane whose size grows with powers of $N$.
In the large $N$ limit, the twisted superpotential becomes a local functional 
\bea
 \wt \cW (\rho( t ), v_a( t ) ; \Delta) \, ,
\eea
of the eigenvalue density $\rho(t)$ and of a set of functions $v_a(t)$ that characterize the gauge holonomies for the $a^{\text{th}}$ group.
This expressions has to be extremized with respect to $\rho(t)$ and $v_a(t)$, and the resulting distribution will then be used to evaluate \eqref{indexBE}.

It was proved in \cite{Hosseini:2016tor} that the ``on-shell'' value of the twisted superpotential for this class of theories
is always related to the free energy on $S^3$, at large $N$, via 
\be
 \mathring{\cW} ( \Delta ) \equiv \wt \cW ( u; \Delta ) \big|_{u = u^*} = - \frac{\ii \pi}{2} F_{S^3} ( \bar \Delta ) \, ,
\ee
where $\bar \Delta \equiv \Delta / \pi$ denote the trial R-charges for the chiral fields. The identification between flavor chemical potentials and R-charges is allowed by the fact the large $N$ saddle point solutions exist for $\sum_{I\in W_a} \Delta_I = 2 \pi$, where $W_a$ denotes a generic monomial term in the superpotential, which correctly enforces   the flavor symmetry constraint  $\prod_{I\in W_a} y_I =1$.

It will be important in the following that we can always use a set of constrained variables $\Delta_I$ such that $\wt \cW (\rho( t ), v^a( t ) ; \Delta)$
contains only \emph{homogeneous} functions of $\Delta_I$ and the on-shell superpotential $\mathring{\cW} ( \Delta ) $ is itself a homogeneous function
of degree two of $\Delta_I$ \cite{Benini:2015eyy,Hosseini:2016tor}. For example, for the ABJM theory we have \cite{Benini:2015noa}%
\footnote{\label{WmapV} To compare with \cite[(3.32)]{Hosseini:2016tor}, note that $\wt \cW_{\text{here}} = - \cV_{\text{there}}$.}
\bea
 \label{ADHM:onshell:WABJM}
 \mathring{\cW} (  \Delta )  = - \frac{2 \ii}{3} N^{3/2} \sqrt{2   \Delta_1  \Delta_2  \Delta_3  \Delta_4} \, ,
\eea
where $\sum_{I=1}^4  \Delta_I = 2 \pi$. 

For $\epsilon \ne 0$ it is impossible to perform the summation over $\fm_i$ in \eqref{RTTI:main}.  A Bethe vacua formula exists for rational $\epsilon$ \cite{Closset:2018ghr},  but it seems difficult to use.  We need to find a different route.

\subsection{The factorization route}\label{gangnnamroute}

The refined twisted index  can be written by gluing two holomorphic blocks $B(u ; \Delta | \epsilon)$  according to the formula \cite{Beem:2012mb}
\be
 \label{Zholfac}
 Z_{S^1 \times S^2_\epsilon} ( \Delta , \fs | \epsilon ) =
 \frac{1}{|\mathfrak{W}|} \sum_{ \fm \in \Gamma_\fh} \int \prod_{i = 1}^{\text{rk}(\cG)}\frac{\rd x_i}{2 \pi \ii x_i} \,  B \left( u^{(1)} ;  \Delta^{(1)} | \epsilon^{(1)} \right) B \left( u^{(2)} ; \Delta^{(2)} | \epsilon^{(2)} \right) \, ,
\ee
where\footnote{The different sign between gauge and flavor fugacity is due to our conventions where $\sum_{I\in W_a} \fs_I =2$ for each term $W_a$ in the superpotential. See appendix \ref{app:gangnam} for details.} 
\bea
 \label{A-twisted:var}
 u^{(\sigma)}_i & \equiv u_i + \frac{\epsilon^{(\sigma)}}{2} \fm_i \, , \qquad \qquad \Delta^{(\sigma)}_I = \Delta_I - \frac{\epsilon^{(\sigma)}}{2} \fs_I \, ,\qquad \qquad  \sigma = 1,2,\\
 \epsilon^{(1)} & \equiv \epsilon \, , \hspace{3.7cm} \epsilon^{(2)} \equiv - \epsilon \, .
\eea
If we assume that, in the limit of interest, the gauge fluxes $\fm_i$ can be treated as continuous variables,
we can then think of \eqref{Zholfac} as an integral over the two independent complex variables $u^{(1)}$ and  $u^{(2)}$ and consider a saddle point with respect to them.  
In a slightly different but equivalent context, the explicit  analysis has been performed in  \cite{Choi:2019dfu}  for the $\cN=8$ theory coupled to a fundamental hypermultiplet, which is supposed to flow to ABJM in the infrared.
The analysis was performed in the Cardy limit $(\epsilon \to 0)$ and leads to the factorized result we are going to discuss
and other similar and interesting results and relations among the topologically twisted index, the superconformal index, and the sphere partition function.

The main point behind this route is the asymptotic expansion of  the holomorphic blocks in the limit of small $\epsilon$. In this limit, the holomorphic blocks are singular (see \eg\;\cite[(2.22)]{Beem:2012mb} and \cite[(F.15)]{Closset:2018ghr}) 
\bea
 \label{Bethehol}
 B ( u ; \Delta | \epsilon ) \underset{\epsilon \to 0}{\sim} \exp \left( \frac{\ii}{\epsilon} \wt \cW ( u ; \Delta ) + \ldots \right) \, , 
\eea
where $\wt \cW (u ; \Delta )$ is the effective twisted superpotential of the two-dimensional theory. For a general class of theories, if we first take the large $N$ limit, the asymptotic series in $\epsilon$ truncates to a polynomial and can be compactly written as
\bea
 \label{Bethehol2}
 B ( u^{(\sigma)} ; \Delta^{(\sigma)} | \epsilon^{(\sigma)} )\sim \exp \left( \frac{\ii}{\epsilon^{(\sigma)}} \wt \cW_{\text{hom}} \left( \rho^{(\sigma)}( t ) ,  v_a^{(\sigma)}( t ) ; \Delta^{(\sigma)} \right)  \right) \, , 
\eea
valid up to exponentially small terms in $\epsilon$, generalizing  \cite{Choi:2019dfu}.
In this formula $\rho^{(\sigma)}( t )$ and  $v_a^{(\sigma)}( t )$ are related to the large $N$ distributions of the variables $u^{(\sigma)}$ and $\cW_{\text hom}$ is the large $N$ twisted superpotential discussed in section \ref{Betheroute}.
It is important that $\wt \cW_{\text{hom}}$ is written \emph{homogeneously} in terms of constrained variables satisfying $\sum_{I \in W_a} \Delta_I = 2 \pi$.
The explicit dependence on $\epsilon$ in \eqref{Bethehol2} comes from replacing $\Delta_I$ with $\Delta_I^{(\sigma)}$, which now satisfy
\be
 \sum_{I\in W_a} \Delta_I^{(\sigma)} = 2 \pi - \epsilon^{(\sigma)} \, .
\ee
The analysis leading to \eqref{Bethehol2} for a chiral contribution is explicitly discussed in appendix \ref{app:gangnam}. 

It follows now from \eqref{Zholfac} and \eqref{Bethehol2} that, treating $u^{(\sigma)}$ as independent variables,
the saddle point analysis reduces to two copies of the extremization discussed in section \ref{Betheroute}.
The final large $N$ limit of the index is then given by  
\be\label{ADHMinndexfac00}
 \log Z = \ii \sum_{\sigma = 1}^2 \frac{\mathring \cW ( \Delta^{(\sigma)} )}{\epsilon^{(\sigma)}} \, ,
\ee
where again we should use the homogeneous form of $\mathring \cW$ for this formula to hold.

\subsection{A direct method}

In this paper we will mostly pursue a direct method. The solution for $\epsilon =0$ suggests that the 
$\fm_i$'s are large and we can treat them effectively as  continuous variables.
Then we can study the saddle point approximation of the refined index matrix model \eqref{RTTI:main} with respect to $u_i$ and $\fm_i$.
The method has been efficiently used to study five-dimensional indices \cite{Hosseini:2018uzp,Jain:2021sdp,Hosseini:2021mnn}. 

Consider for example the contribution of a chiral multiplet to the refined twisted index. At finite $N$ it reads%
\footnote{Up to an overall phase, which including all contributions from vectors and bi-fundamental fields,  is given by  $ -\frac{\ii \pi}{2} \left( \sum_{\alpha \in \cG} \left (\alpha ( \fm ) -1\right ) + \sum_I \sum_{\rho_I \in \fR_I} \left( \rho ( \fm ) - \nu ( \fs ) +1 \right) \right)$. 
The sum over roots and weight vanishes for all quivers where, at each node,  the number of ingoing and outgoing arrows is equal.
The remaining term vanishes at large $N$ since $\Tr R ( \fs ) = 0$ for the quivers of interest, see \eqref{long-range2}.}
\be
 \label{logZ:chiral:ini}
 \log Z_{\chi} = \sum_{\rho \in \fR} \sum_{\ell = - \frac{|B_\rho| - 1}{2}}^{\frac{|B_\rho|-1}{2}} \sign (B_\rho)
 \left( \Li_1 \left( e^{\ii ( \rho ( u ) + \nu ( \Delta ) + \ell \epsilon )} \right)
 + \frac{\ii}{2} g_1 ( \rho ( u )+ \nu ( \Delta ) + \ell \epsilon ) \right) ,
\ee
where $B_\rho \equiv \rho ( \fm ) - \nu ( \fs ) + 1$ and $g_1 (u) = u - \pi$.
We take an ansatz where $\fm_i$ scales with $N$ similarly to the eigenvalues $u_i$ and we will replace them with a set of functions $\fn( t)$ and $\fp_a( t )$.
Then, the chiral contribution \eqref{logZ:chiral:ini} and the full integrand of the refined index \eqref{RTTI:main} become a local functional of $\rho(t )$, $v_a( t)$, $\fn( t)$, and $\fp_a( t )$ that we will be able to extremize directly.
The explicit details are given in appendices \ref{app:N3/2} and \ref{app:N5/3}.

One of the advantage of the direct method is that it provides the distribution of magnetic fluxes that dominates the index at large $N$ and that can be useful
for understanding the quantum mechanics obtained by a dimensional reduction on a sphere, whose ground states are supposed to reproduce the entropy
of magnetically charged black holes in AdS$_4$ \cite{Benini:2022}.

Another advantage of the direct method is that it overcomes one of the technical complication of the Bethe route discussed in section \ref{Betheroute},
namely the possible existence of tails in the large $N$ solution.
They appear in theories with bi-fundamental fields if the eigenvalue distribution, which is typically piece-wise continuous in the large $N$ limit, contains regions where
\bea
 \label{tail}
 u_i^{(b)} - u_i^{(a)} + \Delta_{(b,a)}=  \exp \left( - \sqrt{N} Y_{(b, a)} ( t ) \right) , \qquad i=1,\ldots, N \, .
\eea
Exponentially suppressed terms should be generically negligible, but this is not what happens in the Bethe approach.
The reason is the following.
For a bi-fundamental field, the sum over weights splits into
\bea \sum_{\rho \in \fR} = \sum_{i\ne j}^N +\sum_{i=j}^N \, .\eea
At large $N$, it is usual to discard the terms $\sum_{i=j}^N$ because they are suppressed by a power of $N$ compared to $\sum_{i \ne j}^N$.
Indeed this is what happens when we take the large $N$ limit of the effective twisted superpotential $\wt \cW$.
However, when evaluating \eqref{indexBE} on the Bethe vacuum, we encounter
terms like
\bea
 \sum_{i=1}^N \log \left( u_i^{(b)} - u_i^{(a)} + \Delta_{(b,a)} \right) \sim - N^{3/2} \int \rd t \rho (t) Y_{(b,a)}( t ) \, ,
\eea
which contribute to the leading order for theories with $N^{3/2}$ scaling.
Such contributions are crucial to obtain the correct result for the ABJM theory \cite{Benini:2015eyy}.
On the other hand,  $\epsilon \ne 0$ effectively regularizes and suppresses the $i = j$ contribution to the refined index, as manifest from \eqref{logZ:chiral:ini}, 
and we can neglect problems associated with tail regions. More details are given in appendix \ref{app:N3/2}.

\section{Theories with $N^{3/2}$ scaling of the index}\label{sec:4}

In this section we consider a class of quiver Chern-Simons $\cG = \prod_{a = 1}^{|\cG|} \U(N)_a$ gauge theories with
matters in bi-fundamental, adjoint and (anti-)fundamental representations of the gauge group.
We further require
\be
 \sum_{a = 1}^{|\cG|} k_a = 0 \, .
\ee
The  theories we are interested in are holographically dual to AdS$_4 \times Y_7$
backgrounds of M-theory where $Y_7$ are seven-dimensional Sasaki-Einstein spaces.
They describe the low-energy dynamics of $N$ coincident M2-branes placed at the tip of the cone $\cC (Y_7)$.
In the M-theory phase $N \gg k_a$ the index scales as $N^{3/2}$ as expected from supergravity.

We consider the following ansatz for the large $N$ saddle point eigenvalue distribution
\be
 \label{main:N^3/2:ansatz}
 u^{(a)}_j = \ii N^{1/2} t_j + v^{(a)}_j \, , \qquad \fm^{(a)}_j = \ii N^{1/2} \fn_j + \fp^{(a)}_j \, .
\ee
Observe that we have deformed the real integer fluxes $\fm_j$ into the complex plane in \eqref{main:N^3/2:ansatz}, anticipating a complex saddle point.
Moreover, the imaginary parts of $u_j^{(a)}$ and $\fm_j^{(a)}$ do \emph{not} depend on the index $a$.
At large $N$, we define the continuous functions
\bea
 t_j & \equiv t ( j / N ) \, , \qquad v_j^{(a)} \equiv v^{(a)} ( j / N ) \, , \\
 \fn_j & \equiv \fn ( j / N ) \, , \qquad \fp_j^{(a)} \equiv \fp^{(a)} ( j / N ) \, ,
\eea
and we introduce the normalized density of eigenvalues
\be
 \rho ( t ) = \frac{1}{N} \frac{\rd j}{\rd t} \, , \qquad \int \rd t \? \rho ( t ) = 1 \, .
\ee

Our method generalizes the one used in \cite{Jafferis:2011zi} for the $S^3$ free-energy. As in \cite{Jafferis:2011zi},  there exist some restrictions on the class of quivers for which it can be successfully used. For each bi-fundamental connecting $a$ and $b$ there must be also a bi-fundamental connecting $b$ and $a$,
and the total number of fundamental and anti-fundamental fields in the quiver must be equal.
For such theories to be devoid of long-range forces\footnote{These are non-local terms in the equations of motion which scale with higher powers of $N$.
To obtain a consistent large $N$ limit with the method presented here, they must cancel.} the quiver must also satisfy \cite{Hosseini:2016tor}
\be\label{long-range}
 \prod_{I \in a} y_I = 1 \, , \qquad 2 + \sum_{I \in a} (\fs_I - 1) = 0 \, ,
\ee
where the $\prod_{I \in a}$ and $\sum_{I \in a}$ are taken over all bi-fundamental fields with one leg in the node $a$.%
\footnote{Adjoint fields are counted twice. The second condition is satisfied for most of the quivers of interest that are obtained by dimensionally reducing 4d quivers associated with D3-branes probing Calabi-Yau singularities,  adding Chern-Simons terms  and flavoring with (anti)-fundamentals. See \cite{Hosseini:2016tor,Hosseini:2016ume} for details and examples. For a quiver with a 4d parent, the condition is equivalent to the absence of anomalies for the R-symmetry.} If we sum over all the nodes we also obtain the following constraint
\bea \label{long-range2}
|\cG|+ \sum_{I (\text{bi-fund)}} (\fs_I  -1) = 0 \, .\eea
The above equation is equivalent to $\Tr R = 0$ for any trial R-symmetry. This statement  is valid at large  $N$, where the trace is taken over all the bi-fundamental fermions and gauginos.

\subsection{General rules}

In this section we give the general rules for constructing the large $N$ refined twisted index
of $\cN \geq 2$ quiver gauge theories without long-range forces.
The reader can find the details in Appendix \ref{app:N3/2}, here we only report the final results.

Let us set $w ( t ) = \ii t$.
We define the \emph{equivariant} quantities
\bea\label{equiv}
 w^{(\sigma)} ( t ) & \equiv w ( t ) + \frac{\ii \epsilon^{(\sigma)}}{2} \fn ( t ) \, , \hspace{4.4cm}
v_a^{(\sigma)} ( t )  \equiv  v_a ( t ) + \frac{\epsilon^{(\sigma)}}{2} \fp_a ( t ) \, , \\
 \Delta^{(\sigma)}_{m}  &\equiv \Delta_m - \frac{\epsilon^{(\sigma)}}{2} \fs_m \, , \qquad  \Delta^{(\sigma)}_{I}  \equiv \Delta_I - \frac{\epsilon^{(\sigma)}}{2} \fs_I \, , \qquad  ~ \epsilon^{(1)} \equiv \epsilon \, , \qquad \quad \quad ~ \epsilon^{(2)} \equiv - \epsilon \, ,
\eea
and the closely related ones
\bea
 \bbDelta_I^{(\sigma)} & \equiv \frac{1}{\omega} \left( \Delta_I + \pi ( \omega - 1 ) + \frac{\epsilon^{(\sigma)}}{2} ( 1 - \fs_I ) \right) \, , \qquad
 \tv_{a}^{(\sigma)} ( t ) \equiv \frac{v_a^{(\sigma)} ( t ) }{\omega} \, ,
\eea
with
\be
 \label{main:def:omega:epsilon}
 \omega \equiv \sqrt{1 + \left( \frac{\epsilon}{2 \pi} \right)^2} \, .
\ee
We also define
\be
 \delta v ( t ) \equiv v_b ( t ) - v_a ( t ) \, , \qquad \delta \fp ( t ) \equiv \fp_b ( t ) - \fp_a ( t ) \, .
\ee
We work in an all-orders but perturbative expansion in $\epsilon$. A more refined analysis of the relevant approximations and asymptotic expansions would be needed to understand 
whether there are non-perturbative corrections. We again refers to  Appendix \ref{app:N3/2} for more details.  

\begin{enumerate}
 \item Each gauge group $a$ with CS level $k_a$ contributes
  \be
   \label{main:CS:largeN:3/2}
   \ii N^{3/2} k_a \sum_{\sigma = 1}^2 \frac{1}{\epsilon^{(\sigma)}} \int \rd t \? \rho ( t ) w^{(\sigma)} ( t ) v_a^{(\sigma)} ( t ) \, .
  \ee
 \item A $\U(1)_a$ topological symmetry with chemical potential and magnetic flux $(\Delta_m^{(a)}, \fs_m^{(a)})$ contributes
  \be
   \label{main:top:largeN:3/2}
   \ii N^{3/2} \sum_{\sigma = 1}^2 \frac{\Delta_m^{(\sigma)}}{\epsilon^{(\sigma)}} \int \rd t \? \rho ( t ) w^{(\sigma)} ( t ) \, .
  \ee
 \item Each vector multiplet contributes
  \be
   \label{main:N^3/2:cardy:FINAL:vec}
   \frac{\ii \pi}{12} N^{3/2} \sum_{\sigma = 1}^2
   \int \rd t \? \rho ( t )^2 \? \frac{\epsilon^{(\sigma)} - 2 \pi}{w'^{(\sigma)} ( t )} \, .
  \ee
 \item A pair of bi-fundamental chiral multiplets, one transforming in the $({\bf N},\overline{\bf N})$ representation of $\U(N)_a \times \U(N)_b$ and with chemical potential and magnetic flux $(\Delta_{(a,b)} , \fs_{(a,b)})$
 and the other transforming in the $(\overline{\bf N},{\bf N})$ of $\U(N)_a \times \U(N)_b$ with chemical potential and magnetic flux $( \Delta_{(b,a)}, \fs_{(b,a)} )$, contributes
 \be
  \label{main:N^3/2:cardy:factorized:bifund}
  \ii \omega^3 N^{3/2}
  \sum_{\substack{I = (b , a): + \\ I = (a , b): -}}
  \sum_{\sigma =1}^2
  \frac{1}{\epsilon^{(\sigma)}}
  \int \rd t \? \rho ( t )^2 \?
  \frac{g_3 ( \pm \delta \tv^{(\sigma)} ( t ) + \bbDelta_{I}^{(\sigma)} )}{w'^{(\sigma)}( t )} \, .
 \ee
 Here,  $\delta \tv ( t ) \equiv \tv_b ( t ) -\tv_a ( t )$, and
 \be
  \label{main:def:g:poly}
  g_n ( u ) \equiv \frac{(2 \pi)^n}{n !} B_n \left( \frac{u}{2 \pi} \right) \, , \quad \text{ for } \quad n = 1, 2 , \ldots \, ,
 \ee
 where $B_n ( u )$ denotes the Bernoulli polynomials.
 Note that,
 \be
  \label{main:g123}
  g_1 ( u ) = u - \pi \, , \qquad g_2 ( u ) = \frac{u^2}{2} - \pi u + \frac{\pi^2}{3} \, , \qquad g_3 (u) = \frac{u^3}{6} - \frac{\pi}{2} u^2 + \frac{\pi^2}{3} u \, .
 \ee
 This expression has been derived under the assumption that
 \be
  \pm \delta v^{(\sigma)} ( t ) + \Delta_I^{(\sigma)} \in ( 0 , 2 \pi ) \, .
 \ee
 The solution might contain regions (tails) wherein $\delta v^{(\sigma)} (t )$ is frozen to the constant boundary value $\mp \Delta_I^{(\sigma)}$.
In such regions  the equations obtained from varying the twisted index functional with respect to $\delta v (t)$ and $\delta \fp ( t)$ need \emph{not} hold.\footnote{This is a large $N$ effect. As explained in \cite{Benini:2015eyy}, they hold when including subleading exponential corrections.}
 \item An adjoint chiral multiplet with chemical potential $\Delta_{(a,a)}$ and magnetic flux $\fs_{(a,a)}$, contributes
 \be
  \label{main:N^3/2:cardy:factorized:adj}
  \ii \omega^3 N^{3/2}
  \sum_{\sigma =1}^2
  \frac{g_3 ( \bbDelta_{(a,a)}^{(\sigma)} )}{\epsilon^{(\sigma)}}
  \int \rd t \? \frac{\rho ( t )^2}{w'^{(\sigma)}( t )} \, .
 \ee
 \item A chiral multiplet transforming in the fundamental representation of $\U( N )_a$ and with chemical potential and magnetic flux $(\Delta_a , \fs_a)$, contributes
 \be
  \label{main:N^3/2:fund}
  - \frac{1}{2} N^{3/2} \sum_{\sigma = 1}^2 \int \rd t \? \rho( t )  \left( ( \pi - \Delta_a - v_a ( t ) ) - \frac{\epsilon^{(\sigma)}}{2} ( 1 - \fs_a + \fp_a ( t ) ) \right) \frac{| w^{(\sigma)} |}{\epsilon^{(\sigma)}} \, .
 \ee
 A chiral multiplet in the anti-fundamental representation and with chemical potential and magnetic flux $(\wt \Delta_a , \tilde \fs_a)$, contributes
 \be
  \label{main:N^3/2:anti-fund}
 - \frac{1}{2} N^{3/2} \sum_{\sigma = 1}^2 \int \rd t \? \rho( t ) \left( ( \pi - \wt \Delta_a + v_a ( t ) ) - \frac{\epsilon^{(\sigma)}}{2} ( 1 - \wt \fs_a - \fp_a ( t ) ) \right) \frac{ | w^{(\sigma)} | }{\epsilon^{(\sigma)}} \, .
 \ee
\end{enumerate}

\subsection{ADHM quiver}
\label{sect:ADHM}
The ADHM theory, which arise in the description of the moduli space of instantons \cite{Atiyah:1978ri}, is an $\cN=4$  $\U(N)$ gauge theory with an adjoint hypermultiplet and $r$ fundamental hypermultiplets.
In $\cN=2$ notation, the matter content is described by the quiver diagram 
\bea
\begin{tikzpicture}[font=\footnotesize, scale=0.9]
\begin{scope}[auto,%
  every node/.style={draw, minimum size=0.5cm}, node distance=2cm];
\node[circle]  (UN)  at (0.3,1.7) {$N$};
\node[rectangle, right=of UN] (Ur) {$r$};
\end{scope}
\draw[decoration={markings, mark=at position 0.45 with {\arrow[scale=1.75]{>}}, mark=at position 0.5 with {\arrow[scale=1.75]{>}}, mark=at position 0.55 with {\arrow[scale=1.75]{>}}}, postaction={decorate}, shorten >=0.7pt] (-0,2) arc (30:340:0.75cm);
\draw[draw=black,solid,line width=0.2mm,->]  (UN) to[bend right=30] node[midway,below] {$Q$}node[midway,above] {}  (Ur) ; 
\draw[draw=black,solid,line width=0.2mm,<-]  (UN) to[bend left=30] node[midway,above] {$\wt Q$} node[midway,above] {} (Ur) ;    
\node at (-2.2,1.7) {$\phi_{1,2,3}$};
\end{tikzpicture}
\eea
Here, $\phi_{I}$ with $I=1,2,3$, denotes the adjoint chiral multiplets,
and $Q^i_a$, $\wt Q^a_i$ with $a = 1, \ldots, N$ and $i=1, \ldots, r$ represent the (anti-)fundamental chiral multiplets.
They interact through the superpotential
\be
 W= \wt Q^i_a (\phi_3)^a_{~b} Q^b_i + (\phi_3)^a_{~b}[ \phi_1, \phi_2]^b_{~a} \, .
\ee
With just one fundamental hypermultiplet, $r=1$, the theory is supposed to flow to the ABJM theory with $k=1$ \cite{Aharony:2008ug}. For generic $r$, the theory  can be realized on the world-volume of $N$ M2-branes probing a $\mathbb{C}^2 \times \mathbb{C}^2/\mathbb{Z}_r$ singularity \cite{Porrati:1996xi}.
 
Let us introduce the chemical potentials $(\Delta_I, \Delta, \wt \Delta)$ and magnetic fluxes $(\fs_I, \fs, \wt \fs)$ associated with the fields $(\phi_I, Q, \wt Q)$.
We also denote by $(\Delta_m , \fs_m)$, the chemical potential and magnetic flux corresponding to the topological symmetry associated with  the abelian factor $\U(1)$.
Then,
\bea
 \label{ADHM:constraints}
 \sum_{I = 1}^3 \Delta_I & = 2 \pi \, , \qquad && \Delta + \wt \Delta + \Delta_3 = 2 \pi \, , \\
 \sum_{I = 1}^3 \fs_I & = 2 \, , && \fs + \wt \fs + \fs_3 = 2 \, .
\eea
Here, we use the fact that, for each monomial term $W_a$ in the superpotential the topological twist requires $\sum_{I\in W_a} \fs_I =2$ where the sum is restricted to the fields entering in $W_a$.
The analogous condition for the flavor symmetries is $\prod_{I\in W_a} y_I =1$, which translates into $\sum_{I\in W_a} \Delta_I \in 2 \pi \mathbb{Z}$. As in \cite{Benini:2015noa,Hosseini:2016tor} we will be able to find large $N$ saddle point solutions for $\sum_{I\in W_a} \Delta_I = 2 \pi$. This explains the choices in \eqref{ADHM:constraints}.

Using the rules \eqref{main:top:largeN:3/2}, \eqref{main:N^3/2:cardy:FINAL:vec}, \eqref{main:N^3/2:cardy:factorized:adj}, \eqref{main:N^3/2:fund}, and \eqref{main:N^3/2:anti-fund} the large $N$ refined twisted index reads
\bea
 \label{ADHM:RTTI}
 \frac{\log Z}{N^{3/2}} & = \frac{\ii \pi}{12} \sum_{\sigma = 1}^2 \int \rd t \? \rho ( t )^2 \? \frac{\epsilon^{(\sigma)} - 2 \pi}{w'^{(\sigma)} ( t )}
 + \ii \omega^3 \sum_{I = 1}^3 \sum_{\sigma =1}^2 \frac{g_3 ( \bbDelta_{I}^{(\sigma)} )}{\epsilon^{(\sigma)}} \int \rd t \? \frac{\rho ( t )^2}{w'^{(\sigma)}( t )} \\
 & - \ii \sum_{\sigma = 1}^2 \frac{1}{\epsilon^{(\sigma)}}  \int \rd t \? \rho( t ) \left( \Delta_m^{(\sigma)} - \ii \frac{r}{2} \Delta_3^{(\sigma)} \sign ( w^{(\sigma)} ) \right) w^{(\sigma)} \, ,
\eea
where we used the constraints \eqref{ADHM:constraints}.
The index \eqref{ADHM:RTTI} can be more elegantly rewritten as
\be
 \label{ADHM:Z:factor}
 \log Z ( \rho (t), \fn( t ), \Delta_I , \Delta_m ) = \ii \sum_{\sigma = 1}^2 \frac{\wt \cW_{\text{hom}} \left( \rho (t), w^{(\sigma)} (t), \Delta_I^{(\sigma)}, \Delta_m^{(\sigma)}\right) }{\epsilon^{(\sigma)}} .
\ee
in terms of the effective twisted superpotential for the ADHM theory (see \eg\,\cite[(3.4)]{Hosseini:2016ume})\footnote{In order to recover  the effective twisted superpotential for the ADHM theory as given in \cite{Hosseini:2016ume}, we need to recall that $w(t)=\ii t$. It is convenient to consider an explicit dependence on $w ( t)$ in order to perform the substitutions in \eqref{ADHM:Z:factor}.}
\bea
 \label{ADHM:effective:W}
 \frac{\wt \cW_{\text{hom}} (  \rho (t), w ( t ), \Delta_I , \Delta_m)}{N^{3/2}} & =
 \frac12 \prod_{I = 1}^{3} \Delta_I \int \rd t \? \frac{\rho ( t )^2}{w ' ( t )}
 - \int \rd t \rho ( t ) w ( t ) \left( \Delta_m - \ii \frac{r}{2} \Delta_3 \sign \big( w ( t ) \big) \right) ,
\eea
and we used the equivariant parameters \eqref{equiv}. 
Notice that all terms in \eqref{ADHM:effective:W} have been written as  homogeneous functions of $( \Delta_I, \Delta_m)$ using the constraints \eqref{ADHM:constraints}.
Notice also that the equivariant parameters $\Delta_I^{(\sigma)}$ defined in \eqref{equiv} do \emph{not} satisfy \eqref{ADHM:constraints} and the relation \eqref{ADHM:Z:factor} holds only if  the \emph{homogeneous} form of $\wt\cW$ is used.  

The effective twisted superpotential \eqref{ADHM:effective:W} is at the core of the Bethe route approach used in \cite{Benini:2015noa,Hosseini:2016tor}.
The large $N$ Bethe vacuum that dominates the index is obtained by extremizing \eqref{ADHM:effective:W} with respect to $\rho ( t )$. For completeness and later use,
we report here the result of the extremization. We obtain \cite[Sect.\,3.1.1]{Hosseini:2016ume}
\bea
 \label{ADHM:saddle:W}
 \rho ( t ) & = - \frac{r \Delta_3 | t | + 2 \Delta_m t - 2 \nu}{2 \Delta_1 \Delta_2 \Delta_3} \, , \qquad
 - \frac{2 \nu}{r \Delta_3 - 2 \Delta_m} < t < \frac{2 \nu}{r \Delta_3 + 2 \Delta_m} \, , \\
 \nu & = \sqrt{\frac{r}{2} \Delta_1 \Delta_2 \left( \Delta_3 - \frac{2}{r} \Delta_m \right) \left( \Delta_3 + \frac{2}{r} \Delta_m \right)} \, ,
\eea
where we included the Lagrange multiplier $\nu$ to ensure the normalization of $\rho(t)$.
Plugging back the saddle point configuration \eqref{ADHM:saddle:W} into \eqref{ADHM:effective:W} we then find
\bea
 \label{ADHM:onshell:W}
 \mathring{\cW} ( \Delta_I , \Delta_m ) & \equiv
 \wt \cW_{\text{hom}} ( \rho (t), w ( t ), \Delta_I , \Delta_m ) \Big|_{\eqref{ADHM:saddle:W}}  \\
 & = - \frac{2 \ii}{3} N^{3/2} \nu
 = - \frac{\ii}{3} N^{3/2} \sqrt{2\? r \Delta_1 \Delta_2 \left( \Delta_3 - \frac{2}{r} \Delta_m \right) \left( \Delta_3 + \frac{2}{r} \Delta_m \right)} \, ,
\eea
that is a homogenous function of degree 2 of $(\Delta_I, \Delta_m)$. For $r=1$,
with the change of variables $\wt \Delta_1 = \Delta_1$, $\wt \Delta_2 = \Delta_2$, $\wt \Delta_3 = \frac12 \Delta_3 - \Delta_m$, $\wt \Delta_4 = \frac12 \Delta_3 + \Delta_m$,
\eqref{ADHM:onshell:W} reduces to the  effective twisted superpotential for the ABJM theory \cite{Benini:2015noa}
\bea
 \label{ADHM:onshell:WABJM}
 \mathring{\cW} ( \tilde \Delta_I , \Delta_m )  = - \frac{2 \ii}{3} N^{3/2} \sqrt{2  \wt \Delta_1 \wt \Delta_2 \wt \Delta_3 \wt \Delta_4} \, ,
\eea
where $\sum_{I=1}^4 \wt \Delta_I = 2 \pi$, as expected. 

It was proved in \cite{Hosseini:2016tor} that the ``on-shell'' value of the twisted superpotential for this class of theories
is always related to the free energy on $S^3$, at large $N$, via 
\be
 \mathring{\cW} ( \Delta ) = - \frac{\ii \pi}{2} F_{S^3} ( \bar \Delta ) \, ,
\ee
where $\bar \Delta \equiv \Delta / \pi$ denote the trial R-charges for the chiral fields. 
We will see in the next section that the $S^3$ free energy constitutes the twisted index
in the large $N$ limit, in agreement with the gravitational blocks standpoint proposed in \cite{Hosseini:2019iad}.

\subsubsection{The unrefined case}

Setting $\epsilon = 0$ in \eqref{ADHM:Z:factor} we obtain
\bea
 \label{ADHM:TTI}
 \frac{\log Z}{N^{3/2}} & =
 \int \rd t \? \rho ( t ) \fn ( t ) \left( \Delta_m + \frac{r}{2} \Delta_3 \sign ( t ) \right)
 - \int \rd t \? \rho ( t ) t \left( \fs_m + \frac{r}{2} \fs_3 \sign ( t ) \right) \\
 & - \frac{1}{2} \left( \Delta_1 \Delta_2 \fs_3 + \Delta_1 \Delta_3 \fs_2 + \Delta_2 \Delta_3 \fs_1 \right) \int \rd t \? \rho ( t )^2
 - \frac{1}{2} \Delta_1 \Delta_2 \Delta_3 \int \rd t \? \rho ( t )^2 \fn' ( t ) \, .
\eea
Setting the variations of \eqref{ADHM:TTI} with respect to $\rho( t)$ and $\fn ( t)$ to zero, we find
\be
 \begin{aligned}
 0 & = 2 \left( \mu - \Delta_m \fn ( t ) + t \fs_m \right) + r \sign ( t ) \left( \fs_3 t - \Delta_3 \fn( t ) \right) \\
 & + 2 \rho ( t ) \left( \Delta_1 \Delta_2 \Delta_3 \fn' ( t ) + \Delta_1 \Delta_2 \fs_3 + \Delta_3 \Delta_2 \fs_1 + \Delta_1 \Delta_3 \fs_2 \right) \, , \\
 0 & = \Delta_m + \frac{r}{2} \Delta_3 \sign ( t) + \Delta_1 \Delta_2 \Delta_3 \rho' ( t ) \, ,
 \end{aligned}
\ee
where we introduced the Lagrange multiplier $\mu$ to ensure the normalization of $\rho(t)$.
Thus,
\bea
 \label{ADHM:rho:n:general}
 \rho ( t ) & = -\frac{t ( 2 \Delta_m + r \Delta_3 \sign ( t ) )}{2 \Delta_1 \Delta_2 \Delta_3} + c_1 \, , \\
 \fn ( t ) & = - \frac{1}{2} \left( t \sum_{I=1}^3 \frac{\fs_I}{\Delta_I} - \frac{ t \left( \fs_3 r | t | + 4 \mu + 2 c_1 \left( \Delta_1 \Delta_2 \fs_3 +\Delta_1 \Delta_3 \fs_2 + \Delta_2 \Delta_3 \fs_1 \right) \right)
 + 2 t^2 \fs_m - 2 c_2}{r \Delta_3 | t | - 2 c_1 \Delta_1 \Delta_2 \Delta_3 + 2 t \Delta_m} \right) ,
\eea
where $c_{1,2}$ are constants of integrations. The support of $\rho(t)$ can be easily found by
\be
 \rho ( t_\mp ) = 0 \, \quad \Rightarrow \quad t_\mp = \mp 2 c_1 \frac{\Delta_1 \Delta_2 \Delta_3}{r \Delta_3 \mp 2 \Delta_m} \, .
\ee
Then, the normalization of $\rho(t)$ fixes the value of $c_1$,
\be
 \int_{t_-}^{t_+} \rd t\? \rho ( t ) = 1 \, \quad \Rightarrow \quad c_1 = \frac{1}{\Delta_3} \sqrt{\frac{( r \Delta_3 - 2 \Delta_m ) (r \Delta_3 + 2 \Delta_m )}{2 r \Delta_1 \Delta_2}} \, .
\ee
The constants $\mu$ and $c_2$ are fixed by the requiring that $\fn(t)$ to be regular at the endpoints of the support of $\rho(t)$.
Therefore,
\bea
 \mu & = - \frac{(r \Delta_3 - 2 \Delta_m ) ( r \Delta_3 + 2 \Delta_m ) \left( \Delta_1 \fs_2 + \Delta_2 \fs_1 \right) + 2 \Delta_1 \Delta_2 \left( r^2 \Delta_3 \fs_3 - 4 \Delta_m \fs_m \right)}
 {2 \sqrt{2 r \Delta_1 \Delta_2 ( r \Delta_3 - 2 \Delta_m ) ( r \Delta_3 + 2 \Delta_m )}} \, , \\
 c_2 & = \frac{2 \Delta_1 \Delta_2 \left( \Delta_3 \fs_m - \fs_3 \Delta_m \right)}{r \Delta_3} \, .
\eea
Plugging the constants $c_{1,2}$ and $\mu$ back into \eqref{ADHM:rho:n:general}, we finally find the following large $N$ saddle point solution
\bea
 \label{ADHM:sd:final}
 \rho ( t ) & = \frac{1}{\sqrt{2} \Delta_3} \sqrt{\frac{( r \Delta_3 - 2 \Delta_m ) ( r \Delta_3 + 2 \Delta_m)}{r \Delta_1 \Delta_2}} - \frac{r \Delta_3 | t | + 2 \Delta_m t}{\Delta_1 \Delta_2} \, , \\
 \fn ( t ) & =  \frac{1}{4} \left(\frac{ r \fs_3 + 2 \fs_m}{ r \Delta_3 + 2 \Delta_m} + \frac{r \fs_3 - 2 \fs_m}{r \Delta_3 - 2 \Delta_m} - \frac{2 \fs_1}{\Delta_1} - \frac{2 \fs_2}{\Delta_2} - \frac{2 \fs_3}{\Delta_3}\right) t \\
 & + \frac{( \Delta_3 \fs_m - \fs_3 \Delta_m ) \left( r^2  \Delta_3 | t | + \sqrt{2 r \Delta_1 \Delta_2 ( r \Delta_3 - 2 \Delta_m ) ( r \Delta_3 + 2 \Delta_m )} \right)}
 {r \Delta_3 ( r \Delta_3 - 2 \Delta_m ) ( r \Delta_3 + 2 \Delta_m )} \, .
\eea
Substituting \eqref{ADHM:sd:final} into \eqref{ADHM:TTI} we finally obtain
\be
 \log Z = \frac{2}{3} N^{3/2} \mu = - \frac{N^{3/2}}{3}
 \frac{(r \Delta_3 - 2 \Delta_m ) ( r \Delta_3 + 2 \Delta_m ) \left( \Delta_1 \fs_2 + \Delta_2 \fs_1 \right) + 2 \Delta_1 \Delta_2 \left( r^2 \Delta_3 \fs_3 - 4 \Delta_m \fs_m \right)}
 {2 \sqrt{2 r \Delta_1 \Delta_2 ( r \Delta_3 - 2 \Delta_m ) ( r \Delta_3 + 2 \Delta_m )}} \, ,
\ee
in agreement with \cite[(3.15)]{Hosseini:2016ume}.

\subsubsection{Refined index: $\Delta_m^{(\sigma)} = 0$}

We now turn to evaluate the \emph{refined} twisted index in the branch $\Delta_m = \fs_m = 0$.
Setting the variational derivative of the index \eqref{ADHM:Z:factor} with respect to $\rho(t)$ and $\fn( t )$ to zero,
we find the saddle point
\bea
 \label{ADHM:top=0:saddle}
 \rho ( t ) & = \frac{2 r | t | - \sum_{\sigma = 1}^2 \sqrt{ 2 r \Delta_1^{(\sigma)} \Delta_2^{(\sigma)}}}{\left( \sum_{\sigma = 1}^2 \sqrt{ \Delta_1^{(\sigma)} \Delta_2^{(\sigma)}} \right)^2} \, ,
 \qquad
 - \sum_{\sigma = 1}^2 \sqrt{ \frac{1}{2 r}\Delta_1^{(\sigma)} \Delta_2^{(\sigma)}}  < t < \sum_{\sigma = 1}^2 \sqrt{ \frac{1}{2 r}\Delta_1^{(\sigma)} \Delta_2^{(\sigma)}} \, , \\
 \fn ( t ) & = \frac{2}{\sum_{\sigma = 1}^2 \sqrt{ \Delta_1^{(\sigma)} \Delta_2^{(\sigma)}}} \sum_{\sigma = 1}^2 \frac{\sqrt{ \Delta_1^{(\sigma)} \Delta_2^{(\sigma)} }}{\epsilon^{(\sigma)}} \? t \, , \qquad
 \mu = \sum_{\sigma = 1}^2 \frac{\sqrt{r \Delta_1^{(\sigma)} \Delta_2^{(\sigma)}} \Delta_3^{(\sigma)}}{\sqrt{2} \? \epsilon^{(\sigma)}} \, ,
\eea
where the Lagrange multiplier $\mu$ is included to ensure the normalization of $\rho( t )$.
Plugging back the saddle point \eqref{ADHM:top=0:saddle} into the index \eqref{ADHM:Z:factor} we then obtain
\be
 \log Z = \frac{2}{3} N^{3/2} \mu = \frac{N^{3/2}}{3} \sum_{\sigma = 1}^2 \frac{\sqrt{2 r \Delta_1^{(\sigma)} \Delta_2^{(\sigma)}} \Delta_3^{(\sigma)}}{\epsilon^{(\sigma)}} \, ,
\ee
that can be more elegantly rewritten as
\be
 \log Z = \ii \sum_{\sigma = 1}^2 \frac{\mathring \cW ( \Delta^{(\sigma)} )}{\epsilon^{(\sigma)}} \, ,
\ee
where $\mathring{\cW} ( \Delta )$ is given in \eqref{ADHM:onshell:W}. 

\subsubsection{Refined index: the general case}\label{sec:ADHMgeneral}

One can similarly evaluate the \emph{refined} twisted index for generic values of the chemical potentials.

\begin{figure}
 \centering
 \subfigure[][]{\label{fig:ADHM-rho}
 \includegraphics[width=0.45\linewidth]{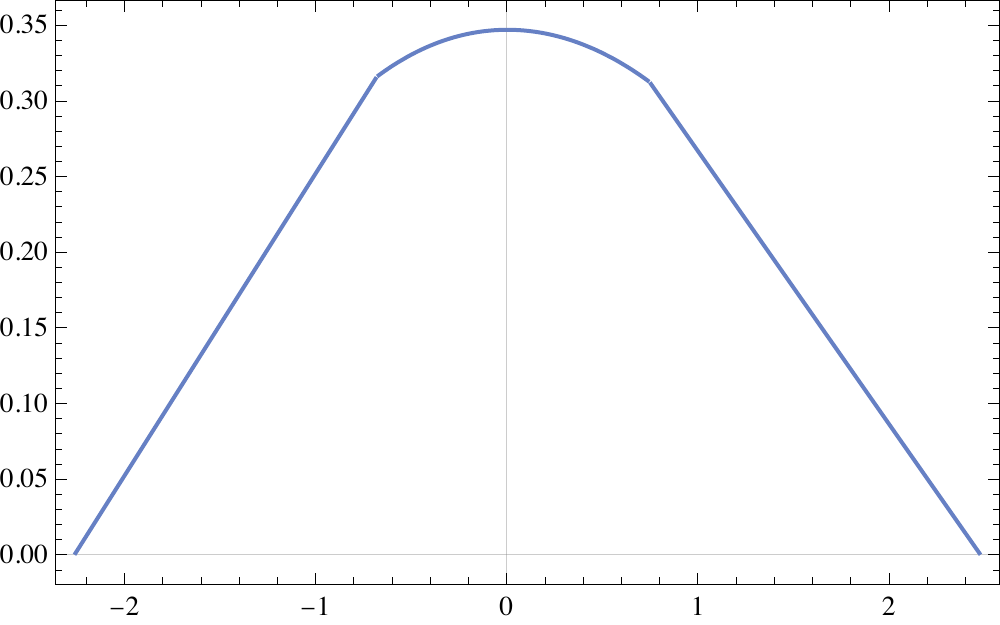}
 \put (0,-1) {\footnotesize $t$}
 \put (-210,140) {\footnotesize $\rho(t)$}
 }\qquad
 \subfigure[][]{\label{fig:ADHM-flux}
 \includegraphics[width=0.45\linewidth]{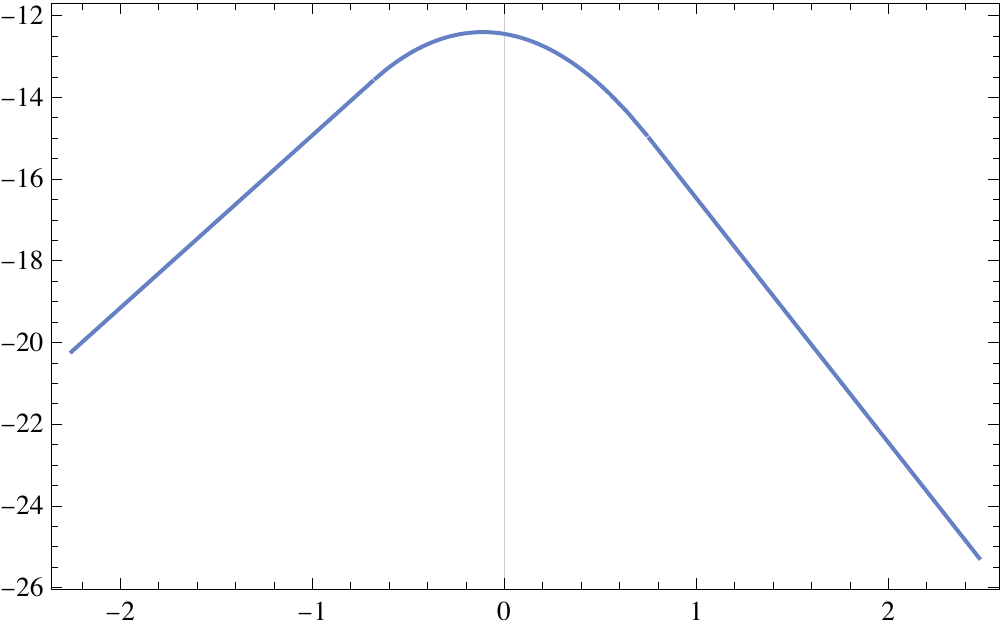}
 \put (0,-1) {\footnotesize $t$}
 \put (-210,140) {\footnotesize $\fn(t)$}
 }
 \caption{Plots of (a) the density of eigenvalues $\rho(t)$ and (b) the function $\fn(t)$ for $r = 2$, $\epsilon = 0.1$,
  $( \Delta_1, \fs_1 ) = ( 2.7 , 1 )$, $( \Delta_2 , \fs_2 ) = ( 1.6, 2)$, $( \Delta_m , \fs_m ) = ( 6.3 - 2 \pi, - 19 )$,
  $\sum_{I = 1}^3 \Delta_I = 2 \pi$, and $\sum_{I = 1}^3 \fs_I = 2$.
 \label{fig:ADHM-rho-flux}}
\end{figure}

We see from the occurrence of $\sign ( w^{(\sigma)} )$ in the functional \eqref{ADHM:RTTI} that the solution is divided into three regions,
delimited by the points where $\rho(t)=0$ and $w^{(\sigma)} (t)= \ii \left(t \pm \frac{\epsilon}{2} n(t) \right )=0$. Schematically, we have:
\begin{center}\label{intervalstADHM}
\begin{tikzpicture}[scale=2]
\draw (-2.,0) -- (2.,0);
\draw (-2,-.05) -- (-2, .05); \draw (-0.7,-.05) -- (-0.7, .05); \draw (0.7,-.05) -- (0.7, .05); \draw (2,-.05) -- (2, .05);
\node [below] at (-2,0) {$t_\ll$}; \node [below] at (-2,-.3) {$\rho=0$};
\node [below] at (-0.7,0) {$t_<$}; \node [below] at (-0.7,-.3) {$t - \frac{\epsilon}{2} \fn ( t ) = 0$};
\node [below] at (0.7,0) {$t_>$}; \node [below] at (0.7,-.3) {$ t + \frac{\epsilon}{2} \fn ( t ) = 0$};
\node [below] at (2,0) {$t_\gg$}; \node [below] at (2,-.3) {$\rho=0$};
\label{intervalsADHM}
\end{tikzpicture}
\end{center}
We assume here that $t - \frac{\epsilon}{2} \fn ( t )$ vanishes before $ t + \frac{\epsilon}{2} \fn ( t )$. This happens for  a region in the space of chemical potentials. The other case is analogous.
A piece-wise continuous solution with these turning points can be explicitly found. The expressions are too long to be reported here  and are given in 
appendix \ref{app:ADHM}. The form of the solution is plotted in figure \ref{fig:ADHM-rho-flux}. 

The refined index is still given by 
\be\label{ADHMinndexfac}
 \log Z = \ii \sum_{\sigma = 1}^2 \frac{\mathring \cW ( \Delta^{(\sigma)} )}{\epsilon^{(\sigma)}} \, ,
\ee
where $\mathring{\cW} ( \Delta )$ is given in \eqref{ADHM:onshell:W}.
We clearly see that the large $N$ index is obtained by $A$-gluing two copies of the twisted superpotential/free energy 
in agreement with the general holographic expectations based on  gravitational blocks \cite{Hosseini:2019iad}.

\subsection{More about factorization}\label{sec:facttrick}
To conclude this example, it is interesting to understand the form of the general solution of section \ref{sec:ADHMgeneral} by connecting the result to the factorization method.
The refined index \eqref{ADHM:Z:factor}  depends explicitly on the distribution of magnetic flux $\fn (t)$ through the variables $ w^{(\sigma)} ( t ) = \ii ( t  + \frac{ \epsilon^{(\sigma)}}{2} \fn ( t ))$. Defining the new  quantites 
\bea
 \ii \? T^{(\sigma)}  =w^{(\sigma)} ( t )  \, , \qquad
 \rho^{(\sigma)} ( T^{(\sigma)} ) = \frac{\ii \? \rho( t )}{ w^{\prime (\sigma)} ( t ) } \, ,
\eea
the index \eqref{ADHM:Z:factor} becomes the sum 
\be
 \label{ADHM:Z:factor2}
 \log Z  = \ii \sum_{\sigma = 1}^2 \frac{\wt \cW_{\text{hom}} \left( \rho^{(\sigma)} ( T^{(\sigma)} ), \Delta_I^{(\sigma)}, \Delta_m^{(\sigma)}\right) }{\epsilon^{(\sigma)}} \, ,
\ee
of the two ADHM twisted superpotentials
\bea
 \label{ADHM:effective:W2}
 \frac{\wt \cW_{\text{hom}} (  \rho (T),  \Delta_I , \Delta_m)}{N^{3/2}} =
 \frac{\ii}2 \prod_{I = 1}^{3} \Delta_I \int \rd T \? \rho ( T )^2
 - \ii \int \rd T \rho ( T ) T \left( \Delta_m + \frac{r}{2} \Delta_3 \sign ( T ) \big) \right) \, .
\eea
The two densities $\rho^{(\sigma)}$ replace the original quantities $\rho( t )$ and $\fn( t )$. They are independent variables  and are correctly normalized
\bea \int \rd T^{(\sigma)} \rho^{(\sigma)}( T^{(\sigma)}) = \int \rd t \rho( t ) =1 \, .\eea
It follows that we can extremize \emph{independently} the two terms in \eqref{ADHM:Z:factor2}, the first with respect to $\rho^{(1)}$ and the second with respect to $\rho^{(2)}$.
Each extremization is equivalent to find the critical point of the ADHM twisted superpotential  with chemical potentials $\Delta^{(1)}$ and $\Delta^{(2)}$, respectively,  and the final result is hence given by
the factorized form \eqref{ADHMinndexfac},  confirming the previous result.

This argument can be also used to  evaluate $\fn( t)$ and $\rho( t)$. The densities $\rho^{(\sigma)}$ are the piece-wise linear functions given in \eqref{ADHM:saddle:W}. $\rho(t)$ and $\fn(t)$ can be determined by the consistency condition 
\bea
 \ii \? \rho( t) = \rho^{(1)}( -\ii w^{(1)}( t) )\? w^{\prime (1)}( t)= \rho^{(2)}( -\ii w^{(2)} ( t))\? w^{\prime (2)}( t)  \, ,
\eea
which follows from the definition of $\rho^{(\sigma)}$. This gives a first order differential equation for $\fn ( t)$ that can be explicitly solved by
\bea\label{match} \int \rd T^{(1)} \rho^{(1)}(T^{(1)}) \Big |_{T^{(1)}= t+ \frac{\epsilon}{2} \fn( t )}=\int \rd T^{(2)} \rho^{(2)}(T^{(2)}) \Big |_{T^{(2)}= t-  \frac{\epsilon}{2}\fn( t )}+ \text{const} \, .\eea
 In the ADHM case this is a quadratic equation for $\fn ( t )$ that should be solved in various intervals. 

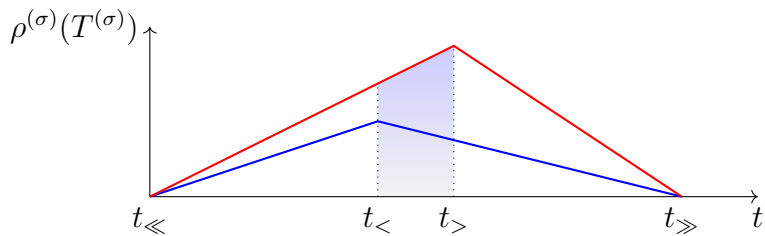
\begin{figure}[h!!!!!]
\centering
\begin{tikzpicture}

\shade[top color=blue!20,bottom color=gray!10] 
      (3,0) -- (4,0) -- (4,2) -- (3,1.5);

\draw[->] (0,0) -- (8,0) node[anchor=north] {$t$};

\draw	(0,0) node[anchor=north] {$t_\ll$}
		(3,0) node[anchor=north] {$t_<$}
		(4,0) node[anchor=north] {$t_>$}
		(7,0) node[anchor=north] {$t_\gg$};

\draw[->] (0,0) -- (0,2.25) node[anchor=east] {$\rho^{(\sigma)}(T^{(\sigma)})$};

\draw[dotted] (3,0) -- (3,1.5);
\draw[dotted] (4,0) -- (4,2);

\draw[thick,blue] (0,0) -- (3,1)-- (7,0);

\draw[thick,red] (0,0) -- (4,2) -- (7,0);

\end{tikzpicture}
\caption{The structure of the intervals in $t$ is obtained by comparing the distributions $\rho^{(1)}$ (in red) and  $\rho^{(2)}$ (in blue), that are the piece-wise linear functions of $T$ given in \eqref{ADHM:saddle:W}.} 
\label{fig:tikz1}
\end{figure}

More explicitly, $ \rho^{(\sigma)}( T^{(\sigma)})$ have support in $\big[T^{(\sigma)}_-,T^{(\sigma)}_+ \big]$, where $T^{(\sigma)}_\pm$ can be read off
from \eqref{ADHM:saddle:W} by replacing $( \Delta_I, \Delta_m)$ with $( \Delta_I^{(\sigma)}, \Delta_m^{(\sigma)})$.  
The structure of intervals  in the variable $t$ arises as follows. Since $ \rho^{(1)}$ and $ \rho^{(2)}$ are piece-wise continuous with junctions at $T^{(1)}=0$ and  $T^{(2)}=0$, which generically correspond to different  values of $t$, there are three  regions for $t$ that are depicted in  figure \ref{fig:tikz1}. We assume that
\bea
 \label{assump:monotonic}
 T^{(1)}( t )  = t +  \frac{ \epsilon}{2} \fn ( t )\, ,
 \qquad T^{(2)} ( t )= t - \frac{ \epsilon}{2} \fn ( t ) \, ,
\eea
are monotonic functions of $t$.
Notice that the assumption \eqref{assump:monotonic} has been used in deriving the rules for the refined index in appendix \ref{app:N3/2}.
We first determine the leftmost point $t_\ll$ as the solution of 
\bea
 T^{(1)}_- = t_\ll +  \frac{ \epsilon}{2} \fn ( t_\ll )\, , \qquad T^{(2)}_- = t_\ll - \frac{ \epsilon}{2} \fn ( t_\ll ) \, .
\eea
A simultaneous solution to these equations can be found by carefully tuning the arbitrary constant in \eqref{match} in the first interval, for example by choosing
\bea
 \label{match2}
 \int^{t+ \frac{\epsilon}{2} \fn( t )}_{T^{(1)}_-} \rd T^{(1)} \rho^{(1)}(T^{(1)}) = \int^{t - \frac{\epsilon}{2} \fn( t )}_{T^{(2)}_-} \rd T^{(2)} \rho^{(2)}(T^{(2)})  \, .
\eea
We can similarly find $t_\gg$ by imposing $T^{(1)}_+ = t_\gg +  \frac{ \epsilon}{2} \fn ( t_\gg )$ and $T^{(2)}_+ = t_\gg - \frac{ \epsilon}{2} \fn ( t_\gg )$ and fixing the constant in the third interval.  Assuming  that $T^{(2)}=0$ occurs before $T^{(1)}=0$ for consistency with the previous section, the point $t_<$
is determined by $T^{(2)} (t_<) = t_< - \frac{ \epsilon}{2} \fn ( t_< )=0$ and  $t_>$ is determined by $T^{(1)} (t_>) = t_> + \frac{ \epsilon}{2} \fn ( t_> )=0$.
There is a remaining arbitrary constant in the relation \eqref{match} in the middle interval 
\bea
 \label{match3}
 \int^{t+ \frac{\epsilon}{2} \fn( t )}_{0} \rd T^{(1)} \rho^{(1)}(T^{(1)}) =\int^{t- \frac{\epsilon}{2} \fn( t )}_{0} \rd T^{(2)} \rho^{(2)}(T^{(2)})  + \text{const} \, ,
\eea
that can be found by imposing that the equation holds at the junctions $t_<$ and $t_>$. It would seem that there are two conditions for one constant, but it is not difficult to see that the two constraints  are equivalent  since the integral of  $ \rho^{(\sigma)}( T^{(\sigma)})$ is normalized to one. By solving explicitly \eqref{match} in the three regions we find a piece-wise continuous function $\rho(t)$   and a piece-wise $C^1$-function $\fn ( t) $ that coincide with the expressions in appendix \ref{app:ADHM}.

\subsection{ABJM quiver}

The ABJM theory \cite{Aharony:2008ug} is a $\U(N)_k\times \U(N)_{-k}$ Chern-Simons gauge theory (the subscripts are the Chern-Simons levels).  In $\cN =2$ notations it is described by the following quiver diagram 
\bea
\begin{tikzpicture}[baseline, font=\footnotesize, scale=0.8]
\begin{scope}[auto,%
  every node/.style={draw, minimum size=0.5cm}, node distance=2cm];
\node[circle] (USp2k) at (-0.1, 0) {$N_k$};
\node[circle, right=of USp2k] (BN)  {$N_{-k}$};
\end{scope}
\draw[decoration={markings, mark=at position 0.9 with {\arrow[scale=1.5]{>}}, mark=at position 0.95 with {\arrow[scale=1.5]{>}}}, postaction={decorate}, shorten >=0.7pt]  (USp2k) to[bend left=40] node[midway,above] {$A_{i}$} node[midway,above] {} (BN) ; 
\draw[decoration={markings, mark=at position 0.1 with {\arrow[scale=1.5]{<}}, mark=at position 0.15 with {\arrow[scale=1.5]{<}}}, postaction={decorate}, shorten >=0.7pt]  (USp2k) to[bend right=40] node[midway,above] {$B_{j}$}node[midway,above] {}  (BN) ;  
\end{tikzpicture}
\eea
with bi-fundamental fields $A_i, \, i=1,2$ and $B_i, \, i=1,2$  transforming in the $(N, \bar N)$ and $(\bar N, N)$ representation of the gauge group, respectively, and  quartic superpotential
\be
 W = \Tr ( A_1 B_1 A_2 B_2 - A_1 B_2 A_2 B_1 ) \, .
\ee
We assign chemical potentials $\Delta_I, \, I=1,\ldots 4$ and fluxes $\fs_I, \, I=1,\ldots 4$ to the fields $A_1,A_2,B_1,B_2$, respectively. Due to the quartic superpotential, they satisfy
\be
 \label{ABJM:const}
 \sum_{I = 1}^4 \Delta_I = 2 \pi \, , \qquad \sum_{I = 1}^4 \fs_I = 2 \, .
\ee
As in \cite{Benini:2015eyy}, the first condition allows to find a large $N$ limit\footnote{The invariance of the superpotential $W$ under flavor symmetries requires $\prod_{I=1}^4 y_I=1$, which is compatible with the more general constraint $\sum_{I = 1}^4 \Delta_I \in 2 \pi \mathbb{Z}$. We know from \cite{Benini:2015eyy} that,  for $\epsilon=0$, a saddle point exists for $\sum_{I = 1}^4 \Delta_I = 2 \pi$.} while the second is a consequence of the topological twist.

Using the rules \eqref{main:CS:largeN:3/2} and \eqref{main:N^3/2:cardy:factorized:bifund}, the large $N$ refined twisted index reads
\bea
 \frac{\log Z}{N^{3/2}} & =
 - \ii N^{3/2} k \sum_{\sigma = 1}^2 \frac{w^{(\sigma)} ( t ) \delta v^{(\sigma)} ( t )}{\epsilon^{(\sigma)}}
 + \frac{\ii \pi}{6} \sum_{\sigma = 1}^2
 \int \rd t \? \rho ( t )^2 \? \frac{\epsilon^{(\sigma)} - 2 \pi}{w'^{(\sigma)} ( t )} \\
 & + \ii \omega^3 \sum_{\substack{I = (3 , 4): + \\ I = (1 , 2): -}} \sum_{\sigma =1}^2 \frac{1}{\epsilon^{(\sigma)}}
 \int \rd t \? \rho ( t )^2 \? \frac{g_3 ( \pm \delta \tv^{(\sigma)} ( t ) + \bbDelta_{I}^{(\sigma)} )}{w'^{(\sigma)}( t )} \, ,
\eea
that can be more elegantly rewritten as
\be
 \label{ABJM:Z:factor}
 \log Z ( \rho (t), w^{(\sigma)} (t), \Delta_I^{(\sigma)}) = \ii \sum_{\sigma = 1}^2 \frac{\wt \cW_{\text{hom}} \left( \rho (t), w^{(\sigma)}( t ), \delta v^{(\sigma)} ( t ) , \Delta^{(\sigma)}_I \right) }{\epsilon^{(\sigma)}} .
\ee
Here, $\wt \cW_{\text{hom}} (  \rho (t), w( t ), \delta v ( t ) , \Delta_I )$, with $w( t )=\ii t$, is the effective twisted superpotential, see \eg\,\cite[(3.28)]{Hosseini:2016tor},
\bea
 \label{ABJM:effective:W}
 \frac{\wt \cW_{\text{hom}} (  \rho (t), w ( t ), \delta v ( t ) , \Delta_I )}{N^{3/2}} & = - k \int \rd t \? \rho ( t ) w ( t ) \delta v ( t )
 + \sum_{\substack{I = (3 , 4): + \\ I = (1 , 2): -}} \int \rd t \? \rho ( t )^2 \frac{g_3 ( \pm \delta v ( t ) + \Delta_I )}{w' ( t )} \, ,
\eea
and we used again the $A$-gluing parameterization
\bea
 \label{N^3/2:ABJM:cov}
 w^{(\sigma)} ( t ) & \equiv w ( t ) + \ii \frac{\epsilon^{(\sigma)}}{2} \fn ( t ) \, , \qquad \delta v^{(\sigma)} \equiv \delta v ( t ) + \frac{\epsilon^{(\sigma)}}{2} \delta \fp ( t ) \, , \qquad \Delta^{(\sigma)} = \Delta_I - \frac{\epsilon^{(\sigma)}}{2} \fs_I \, , \\
 \epsilon^{(1)} & \equiv \epsilon \, , \hspace{3.7cm} \epsilon^{(2)} \equiv - \epsilon \, .
\eea
Importantly, the subscript \emph{hom} in  \eqref{ABJM:effective:W} indicates that the polynomial $\sum_{\substack{I }} g_3 ( \pm \delta v ( t ) + \Delta_I )$ in \eqref{ABJM:effective:W} must be written as a \emph{homogeneous} function of $\Delta_I$ using the constraint \eqref{ABJM:const}.  Only if this is done, \eqref{ABJM:Z:factor} holds.%
\footnote{Notice that $ \sum_{I = 1}^4 \Delta^{(\sigma)}_I \ne 2 \pi$, so to what extent the relation \eqref{ABJM:const} has been used before substituting $\Delta_I \rightarrow \Delta_I^{(\sigma)}$ is important.}

The more efficient way of extremizing \eqref{ABJM:Z:factor} for generic $\epsilon$ is to use the factorization trick discussed in section \ref{sec:facttrick}. Defining the new  quantites 
\bea
 \label{mapp}
 \ii \? T^{(\sigma)}  =w^{(\sigma)} ( t )  \, , \qquad
 \rho^{(\sigma)} ( T^{(\sigma)} ) = \frac{\ii \? \rho( t )}{ w^{\prime (\sigma)} ( t ) } \, ,
\eea
the index \eqref{ABJM:Z:factor} becomes the sum 
\be
 \label{ABJM:Z:factor2}
 \log Z  = \ii \sum_{\sigma = 1}^2 \frac{\wt \cW_{\text{hom}} \left( \rho^{(\sigma)} ( T^{(\sigma)} ), \delta v^{(\sigma)} ( T^{(\sigma)} ), \Delta_I^{(\sigma)} \right) }{\epsilon^{(\sigma)}} \, ,
\ee
of the two ABJM twisted superpotentials
\bea
 \label{ABJM:effective:W2}
 \frac{\wt \cW_{\text{hom}} (  \rho ( T ), \delta v ( T ) , \Delta_I )}{N^{3/2}} & = - k \? \ii \int \rd t \? \rho ( T ) T \delta v ( T )
 - \ii \sum_{\substack{I = (3 , 4): + \\ I = (1 , 2): -}} \int \rd T \? \rho ( T )^2 g_3 ( \pm \delta v ( T ) + \Delta_I )\, .
\eea
As in section \ref{sec:facttrick}  we can extremize \emph{independently} the two terms in \eqref{ABJM:Z:factor2}. The extremization of \eqref{ABJM:effective:W2} was performed in
\cite{Benini:2015eyy} finding the following distribution for $\sum_I \Delta_I = 2 \pi$ and $\Delta_1\leq \Delta_2$, $\Delta_3\leq \Delta_4$.
We have a central region where
\be
 \label{ABJM1}
 \begin{aligned}
 \rho (T ) & = \frac{\nu \sum_{I = 1}^4 \Delta_I + k T(\Delta_3 \Delta_4 - \Delta_1 \Delta_2)}{(\Delta_1 + \Delta_3)(\Delta_2 + \Delta_3)(\Delta_1 + \Delta_4)(\Delta_2 + \Delta_4)} \, , \\[.5em]
 \delta v ( T ) & = \frac{\nu(\Delta_1 \Delta_2 - \Delta_3 \Delta_4) + k T \sum_{I < J < K} \Delta_I \Delta_J \Delta_K }{\nu \sum_{I = 1}^4 \Delta_I + k T ( \Delta_3 \Delta_4 - \Delta_1 \Delta_2) } \, ,
\end{aligned}
\qquad\qquad -\frac{\nu}{k \Delta_4}   < T < \frac{\nu}{k \Delta_2} \, ,
\ee
with $\nu = \sqrt{ 2 k \Delta_1 \Delta_2 \Delta_3 \Delta_4}$.
There is a left tail where $\delta v$ is frozen to the value $-\Delta_3$,
\be\label{ABJM2}
\rho  ( T ) = \frac{\nu + k T\Delta_3}{(\Delta_1 + \Delta_3)(\Delta_2 + \Delta_3)(\Delta_4 - \Delta_3)} \, , \qquad  -\frac{\nu}{k \Delta_3}  < T <-\frac{\nu}{k \Delta_4}  \, ,
\ee
and a right tail with $\delta v = \Delta_1$,
\be\label{ABJM3}
\rho (T ) = \frac{\nu - k T \Delta_1}{(\Delta_1 + \Delta_3)(\Delta_1 + \Delta_4)(\Delta_2 - \Delta_1)} \, , \qquad \frac{\nu}{k \Delta_2}  < T < \frac{\nu}{k \Delta_1} \, .
\ee
The on-shell twisted superpotential is given by
\bea
 \label{onshell:WABJM}
 \mathring{\cW} (  \Delta_I )  = - \frac{2 \ii}{3} N^{3/2} \sqrt{2  k \Delta_1  \Delta_2  \Delta_3  \Delta_4} \, .
\eea 
We thus find for the refined index
\be\label{refinedABJM}
 \log Z = \ii \sum_{\sigma = 1}^2 \frac{\mathring \cW ( \Delta^{(\sigma)} )}{\epsilon^{(\sigma)}} \, ,
\ee
in agreement with the general holographic expectations based on  gravitational blocks \cite{Hosseini:2019iad}.

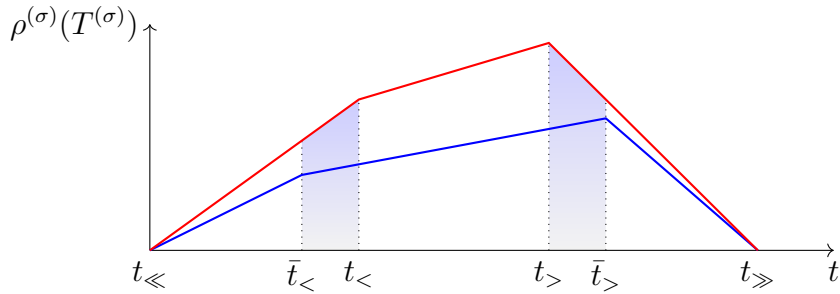
\begin{figure}[h!!!!!]
\centering
\begin{tikzpicture}

\shade[top color=blue!20,bottom color=gray!10] 
      (2,0) -- (2.75,0) -- (2.75,2) -- (2,1.45);

\shade[top color=blue!20,bottom color=gray!10] 
      (5.25,0) -- (6,0) -- (6,2) -- (5.25,2.75);

\draw[->] (0,0) -- (9,0) node[anchor=north] {$t$};

\draw	(0,0) node[anchor=north] {$t_\ll$}
		(2,0) node[anchor=north] {$\bar t_<$}
		(2.75,0) node[anchor=north] {$t_<$}
		(5.25,0) node[anchor=north] {$t_>$}
		(6,0) node[anchor=north] {$\bar t_>$}
		(8,0) node[anchor=north] {$t_\gg$};

\draw[->] (0,0) -- (0,3) node[anchor=east] {$\rho^{(\sigma)}( T^{(\sigma)} )$};

\draw[dotted] (2,0) -- (2,1.45);
\draw[dotted] (2.75,0) -- (2.75,2);
\draw[dotted] (5.25,0) -- (5.25,2.75);
\draw[dotted] (6,0) -- (6,2);

\draw[thick,blue] (0,0) -- (2,1) -- (6,1.75) -- (8,0);

\draw[thick,red] (0,0) -- (2.75,2) -- (5.25,2.75) -- (8,0);

\end{tikzpicture}
\caption{The structure of the intervals in $t$ is obtained by comparing the distributions $\rho^{(1)}$ (in red) and  $\rho^{(2)}$ (in blue), that are the piece-wise linear functions of $T$ given in \eqref{ABJM1}, \eqref{ABJM2}, and \eqref{ABJM3}.} 
\label{fig:tikz2}
\end{figure}

As in section \ref{sec:facttrick}, the quantities $\rho( t )$, $\delta v( t)$, $\fn ( t)$, and $\fp( t )$ can be explicitly obtained by using the mapping \eqref{mapp}. In particular, as clear from figure  
\ref{fig:tikz2}, the solution will be divided into five regions. In each region,  \eqref{mapp} provides a differential equation for $\fn( t )$. Imposing that the distribution is piece-wise continuous, we find a unique solution.

In appendix \ref{app:ABJM}, for reference, we give the full explicit solution  for $\epsilon =0$, which can be also find by a direct  extremization of \eqref{ABJM:Z:factor}. Notice that for $\epsilon=0$ the five segments collapse to three and the functions $\rho( t )$  and $\delta v ( t )$ are the same as in \eqref{ABJM1}, \eqref{ABJM2}, and \eqref{ABJM3}. The distribution of magnetic fluxes $\fn ( t )$ is given explicitly  in the appendix.

\paragraph*{Comparison to the entropy of rotating AdS$_4 \times S^7$ black holes.}
Here we simply notice that the result \eqref{refinedABJM} correctly reproduced the entropy of the rotating AdS$_4 \times S^7$ black holes found in \cite{Hristov:2018spe}.
This was explicitly checked in \cite[Sect.\,2]{Hosseini:2019iad}, where a general formalism based on gravitational blocks was proposed.  
The factorization in \eqref{refinedABJM} is quantum field theory analog of the gravitational block factorization.

\section{Theories with $N^{5/3}$ scaling of the index}\label{sec:5}

We now consider Chern-Simons  $\cG = \prod_{a = 1}^{|\cG|} \U(N)_a$ gauge theories with
matters in bi-fundamental and adjoint  representations of the gauge group
that are holographically dual to AdS$_4$
backgrounds in massive type IIA. We assume 
\be
 k_{\text{CS}} \equiv \sum_{a = 1}^{|\cG|} k_a \neq 0 \, ,
\ee
that corresponds to turning on the Romans mass $F_0$ in the dual type IIA supergravity \cite{Gaiotto:2009mv}.
In the limit $N\gg k_a$ we can find a class of theories whose free-energy and indices scale as  $N^{5/3}$ \cite{Aharony:2010af,Guarino:2015jca,Fluder:2015eoa}.
They are obtained by dimensionally reducing 4d quivers associated with D3-branes probing Calabi-Yau singularities  and adding Chern-Simons terms.

Following \cite{Jafferis:2011zi,Hosseini:2016tor}, we consider the following ansatz for the large $N$ saddle point eigenvalue distribution
\be
 \label{app:N^5/3:ansatz}
 u^{(a)} (t) = N^{1/3} ( \ii t + v(t) ) \, , \qquad \fm^{(a)} ( t ) = \ii N^{1/3} \fn(t) \, .
\ee
Notice that the distribution is the same for all groups. We still need to satisfy \eqref{long-range} (or a milder condition) but now the saddle-point exists also for chiral quivers  \cite{Hosseini:2016tor}.

\subsection{General rules}

Let us set $w ( t ) = \ii t + v ( t )$. We define the \emph{equivariant} quantities
\bea\label{cov}
w^{(\sigma)} ( t ) & \equiv w ( t ) + \ii \frac{\epsilon^{(\sigma)}}{2} \fn ( t ) \, , && \Delta_I^{(\sigma)} = \Delta_I - \frac{\epsilon^{(\sigma)}}{2} \fs_I \, , \\
 \epsilon^{(1)} & = \epsilon \, , \hspace{3cm} && \epsilon^{(2)} = - \epsilon \, ,
\eea
and the closely related quantity
\bea
\bbDelta_I^{(\sigma)} \equiv \frac{1}{\omega} \left( \Delta_I + \pi ( \omega - 1 ) + \frac{\epsilon^{(\sigma)}}{2} ( 1 - \fs_I ) \right) ,
\eea
with $\omega$ as before, see \eqref{main:def:omega:epsilon}.

\begin{enumerate}
 \item Each gauge group $a$ with CS level $k_a$ contributes
  \be
   \label{main:CS:largeN:5/3}
   - k_a N^{5/3} \int \rd t \? \rho ( t ) \fn ( t ) w ( t ) \, .
  \ee
 \item Each vector multiplet contributes
  \be
   \label{main:N^5/3:cardy:FINAL:vec}
   \frac{\ii \pi}{12} N^{5/3} \sum_{\sigma = 1}^2
   \int \rd t \? \rho ( t )^2 \? \frac{\epsilon^{(\sigma)} - 2 \pi}{w'^{(\sigma)} ( t )} \, .
  \ee
 \item A single bi-fundamental chiral multiplet transforming in a representation $({\bf N},\overline{\bf N})$ of $\U(N)_a \times \U(N)_b$
 and with chemical potential and magnetic flux $(\Delta_{(a,b)}, \fs_{(a,b)})$ contributes
  \be
   \label{main:cardy:factorized:chi}
   \ii \omega^3 N^{5/3} \sum_{\sigma = 1}^2 \frac{g_3 \big( \bbDelta_{(a,b)}^{(\sigma)} \big)}{\epsilon^{(\sigma)}} \int \rd t \? \frac{\rho ( t )^2}{w'^{(\sigma)}( t )} \, .
  \ee
\end{enumerate}

\subsection{Large $N$ twisted index: $\epsilon = 0$ case}

Let us consider a generic three-dimensional $\cN = 2$ Chern-Simons-matter quiver theory, with $|\cG|$
$\U(N)$ gauge nodes and some number of bi-fundamental and adjoint chiral multiplets of the type considered in  \cite{Guarino:2015jca,Fluder:2015eoa}.
Using \eqref{main:CS:largeN:5/3}, \eqref{main:N^5/3:cardy:FINAL:vec}, and \eqref{main:cardy:factorized:chi} the large $N$ twisted index for $\epsilon=0$ can be written as
\bea
 \label{logZ:noepsilon:generic}
 \frac{\log Z}{N^{5/3} } & = - k_{\text{CS}} \int \rd t \? \rho ( t ) \fn ( t ) \left( \ii t + v ( t ) \right)
 - \sum_I g_3 (\Delta_I) \int \rd t \frac{\rho ( t )^2 \fn' ( t )}{\left( 1 - \ii v' ( t ) \right)^2} \\
 & - \bigg( |\cG| \frac{\pi^2}{3} +\sum_{I } ( \fs_I - 1 ) g_2 ( \Delta_I ) \bigg) \int \rd t \frac{\rho ( t )^2 }{1-i v'(t)}
 + \mu \left( \int \rd t \rho (t) - 1 \right) ,
\eea
where $k_{\text{CS}} \equiv \sum_{a = 1}^{|\cG|} k_a$ and we introduced the Lagrange multiplier $\mu$ to ensure the normalization of the density of eigenvalues.
Here, $\Delta_I$ and $\fs_I$ denote the chemical potentials and magnetic charges for the flavor group of the theory. As before, they satisfy
\bea\label{supo} \sum_{I \in W_a} \Delta_I = 2 \pi \, ,\qquad\qquad \sum_{I \in W_a} \fs_I = 2 \, ,\eea
for each superpotential terms $W_a$.
Extremizing \eqref{logZ:noepsilon:generic} with respect to the continuous functions $\rho ( t )$, $\fn ( t)$, and $v( t )$
we find the following general solution
\bea
 \rho ( t ) & = \frac{3^{1/6}}{2} \left( \frac{k_{\text{CS}}}{\sum_I g_3 ( \Delta_I )} \right)^{1/3}
 - \frac{2}{3^{3/2}} \frac{k_{\text{CS}}}{\sum_I g_3 (\Delta_I)} \? t^2 \, , \\
 \fn ( t ) & = - \frac{1}{3} \left( 1 + \frac{\ii}{\sqrt{3}} \right) \frac{|\cG| \frac{\pi^2}{3} +\sum_{I } ( \fs_I - 1 ) g_2 ( \Delta_I )}{\sum_I g_3 (\Delta_I)} \? t \, , \\
 v (t ) & = - \frac{1}{\sqrt{3}} \? t \, , \\
 \mu & = \frac{3^{1/6}}{2} \left( 1 - \frac{\ii}{\sqrt{3}} \right) \left( \frac{k_{\text{CS}}}{\sum_I g_3 ( \Delta_I )} \right)^{1/3}
 \bigg( |\cG| \frac{\pi^2}{3} +\sum_{I } ( \fs_I - 1 ) g_2 ( \Delta_I ) \bigg) \, , \\
 t_{\pm} & = \pm \frac{3^{5/6}}{2} \left( \frac{\sum_I g_3 ( \Delta_I )}{k_{\text{CS}}} \right)^{1/3} .
\eea
Using the above solution, we obtain
\be
 \log Z = - \frac{3}{5} N^{5/3} \mu =
 - N^{5/3} \? \frac{3^{7/6}}{10} \left( 1 - \frac{\ii}{\sqrt{3}} \right) \left( \frac{k_{\text{CS}}}{\sum_I g_3 ( \Delta_I )} \right)^{1/3}
 \bigg( |\cG| \frac{\pi^2}{3} +\sum_{I } ( \fs_I - 1 ) g_2 ( \Delta_I ) \bigg)  \, ,
\ee
that matches the result for the large $N$ twisted index evaluated via the Bethe ansatz approach \cite{Hosseini:2017fjo,Benini:2017oxt,Azzurli:2017kxo}.\footnote{The index can be written as a homogeneous function of $\Delta_I$ using \eqref{supo}.}
\subsection{Large $N$ twisted index with refinement}
Given the expressions \eqref{main:CS:largeN:5/3},  \eqref{main:N^5/3:cardy:FINAL:vec}, and  \eqref{main:cardy:factorized:chi},
the \emph{refined} twisted index at large $N$ can be written as
\be
 \label{refined:logZ:index}
 \log Z ( \rho (t), \fn, \Delta) = \ii \sum_{\sigma = 1}^2 \frac{\wt \cW_{\text{hom}} \left( \rho (t), w^{(\sigma)} ( t ) , \Delta^{(\sigma)} \right)}{\epsilon^{(\sigma)}} \, ,
\ee
where $\wt \cW_{\text{hom}} ( \rho (t), w ( t ), \Delta)$ is the effective twisted superpotential, see \eg\,\cite[(2.4)]{Hosseini:2017fjo},
\bea
 \label{effective:W}
 \frac{\wt \cW_{\text{hom}} ( \rho (t), w (t), \Delta)}{N^{5/3}} & = \frac12 k_{\text{CS}} \int \rd t \? \rho ( t ) w ( t )^2
 + \sum_I g_3 ( \Delta_I ) \int \rd t \? \frac{\rho(t)^2}{w' ( t )} \, ,
\eea
written homogeneously in $\Delta_I$, and we used the $A$-gluing parameterization \eqref{cov}.
One can show that 
\be
 G_3 ( \Delta ) \equiv \sum_I g_3 ( \Delta_I )  \bigg|_\text{hom} = \frac{1}{3!} \sum_{I, J , K } c_{I J K} \Delta_I \Delta_J \Delta_K \, ,
\ee
is proportional to the trial $a$-charge of the \emph{parent} four-dimensional theory \cite{Fluder:2015eoa,Hosseini:2017fjo}, which can be written in a homogeneous
form using  the (rescaled) 't\,Hooft anomaly coefficients $c_{IJK}$.%
\footnote{For a four-dimensional toric quiver associated with D3-branes at a Calabi-Yau conical singularity, $c_{I J K} = | \det(v_I, v_J, v_K)| / 2$, where $v_I \in \bZ^3$ are the integer vectors defining the toric diagram \cite{Benvenuti:2006xg}.}
For example, for the $\cN = 8$ super Yang-Mills at Chern-Simons level $k$ \cite{Guarino:2015jca}
\be
G_3 (\Delta ) = \frac12 \Delta_1 \Delta_2 \Delta_3 \, .
\ee
Then, setting to zero the variational derivatives of \eqref{refined:logZ:index} with respect to $\rho ( t)$ and $w^{(\sigma)} ( t)$ yields
\bea
 \label{refined:variations}
 0 & = - 2 \ii \mu + k_{\text{CS}} \sum_{\sigma = 1}^2 \frac{w^{(\sigma)} ( t )^2}{\epsilon^{(\sigma)}}
 + 4 \rho (t) \sum_{\sigma = 1}^2 \frac{1}{\epsilon^{(\sigma)}} \frac{G_3 ( \Delta^{(\sigma)} )}{w'^{(\sigma)} ( t )} \, , \\
 0 & = k_{\text{CS}} \? w^{(\sigma)} ( t ) w'^{(\sigma)} ( t ) + 2 G_3 ( \Delta^{(\sigma)} ) \left( \frac{\rho' ( t )}{w'^{(\sigma)} ( t )} - \frac{\rho ( t ) w''^{(\sigma)} ( t )}{w'^{(\sigma)} ( t )^2} \right) , \quad \text{ for } \quad \sigma = 1, 2 \, .
\eea
Observe that $\frac{\delta \log Z}{\delta w^{(\sigma)} ( t )} = 0$ can be rewritten as
\be
 0 = \frac{\rd}{\rd t} \bigg( \frac{1}{2} k_{\text{CS}} \? w^{(\sigma)} ( t )^2 + 2 G_3 ( \Delta^{(\sigma)} ) \? \frac{\rho (t)}{w'^{(\sigma)} ( t )} + c^{(\sigma)} \bigg) \, , \quad \text{ for } \quad \sigma = 1, 2 \, ,
\ee
where $c^{(\sigma)}$ are constants of integrations to be determined later.
This together with $\frac{\delta \log Z}{\delta \rho ( t )} = 0$ allows us to fix the Lagrange multiplier $\mu$ as
\be
 \mu = \ii \sum_{\sigma = 1}^2 \frac{c^{(\sigma)}}{\epsilon^{(\sigma)}} \, .
\ee
Given the above equations it is now straightforward to check that
\bea
 \label{saddle:refined:mIIA}
 \rho ( t ) & = \frac{3^{1/6} k_{\text{CS}}^{1/3}}
 {\sum_{\sigma = 1}^2 G_3 (\Delta^{(\sigma)})^{1/3}}
 -\frac{16 \? k_{\text{CS}}}{3^{3/2} \left( \sum_{\sigma = 1}^2 G_3 (\Delta^{(\sigma)})^{1/3} \right)^3} \? t^2\, , \\
 w^{(\sigma)} ( t ) & = 2 \ii \left( 1 + \frac{\ii}{\sqrt{3}} \right)
 \frac{G_3 ( \Delta^{(\sigma)})^{1/3}}
 {\sum_{\sigma = 1}^2 G_3 (\Delta^{(\sigma)})^{1/3}} \? t \, , \\
 \mu & = - \frac{3^{7/6}}{4} \left( 1 - \frac{\ii}{\sqrt{3}} \right) k_{\text{CS}}^{1/3}
 \sum_{\sigma = 1}^2 \frac{G_3 (\Delta^{(\sigma)})^{2/3}}{\epsilon^{(\sigma)}} \, , \\
  t_{\pm} & = \pm \frac{3^{5/6}}{4 k_{\text{CS}}^{1/3}} \sum_{\sigma = 1}^2 G_3 (\Delta^{(\sigma)})^{1/3} \, , \\
  c^{(\sigma)} & = \ii \frac{3^{7/6}}{4} \left( 1 - \frac{\ii}{\sqrt{3}} \right) k_{\text{CS}}^{1/3} \?G_3 ( \Delta^{(\sigma)} )^{2/3} \, ,
\eea
satisfies \eqref{refined:variations}. Note that,
\be
 w ( t ) \overset{\eqref{cov}}{=} \frac12 \sum_{\sigma = 1}^2 w^{(\sigma)} ( t ) \overset{\eqref{saddle:refined:mIIA}}{=} \ii \left( 1 + \frac{\ii}{\sqrt{3}} \right) t
 \quad \Rightarrow \quad v ( t ) = - \frac{1}{\sqrt{3}} \? t \, .
\ee
Moreover,
\be
 \fn ( t ) = - \ii \sum_{\sigma = 1}^2 \frac{w^{(\sigma)} ( t )}{\epsilon^{(\sigma)}} \overset{\eqref{saddle:refined:mIIA}}{=} 2 \left( 1 + \frac{\ii}{\sqrt{3}} \right)
 \frac{1}{\sum_{\sigma = 1}^2 G_3 (\Delta^{(\sigma)})^{1/3}} \sum_{\sigma = 1}^2 \frac{G_3 (\Delta^{(\sigma)})^{1/3}}{\epsilon^{(\sigma)}} \? t \, .
\ee
Substituting back the solution \eqref{saddle:refined:mIIA} into \eqref{refined:logZ:index}, we find that the large $N$ refined index takes the following factorized form
\be
 \label{mIIA:logZ:FNL:explicit}
 \log Z = - \frac{3}{5} N^{5/3} \mu =
 N^{5/3} k_{\text{CS}}^{1/3} \?
 \frac{9 \times 3^{1/6}}{20}
 \left( 1 - \frac{\ii}{\sqrt{3}} \right)
 \sum_{\sigma = 1}^2 \frac{G_3 (\Delta^{(\sigma)})^{2/3}}{\epsilon^{(\sigma)}} \, ,
\ee
that can be more elegantly rewritten as
\be
 \log Z = \ii \sum_{\sigma = 1}^2 \frac{\mathring \cW ( \Delta^{(\sigma)} )}{\epsilon^{(\sigma)}} \, ,
\ee
where $\mathring{\cW} ( \Delta )$ is the on-shell value of the effective twisted superpotential \eqref{effective:W}, see \eg\;\cite[(A.5)]{Azzurli:2017kxo},
\be
 \mathring \cW ( \Delta )
 = - \ii N^{5/3} k_{\text{CS}}^{1/3} \, \frac{9 \times 3^{1/6}}{20} \left( 1 - \frac{\ii}{\sqrt{3}} \right) G_3 (\Delta)^{2/3} \, .
\ee
This expression was derived in \cite{Hosseini:2017fjo,Benini:2017oxt,Azzurli:2017kxo} with the Bethe approach and it is proportional to $S^3$ free energy at large $N$ \cite{Guarino:2015jca,Fluder:2015eoa}.
Once again we see that the result is in precise agreement with the general holographic expectations based on gravitational blocks \cite{Hosseini:2019iad}.


\section*{Acknowledgements}

SMH is supported in part by the STFC Consolidated Grant ST/T000791/1.
AZ is partially supported by the INFN, and the MIUR-PRIN contract 2017CC72MK003.

\appendix

\section{Derivation of general rules for quivers with $N^{3/2}$ scaling of the index}
\label{app:N3/2}

We consider the following ansatz for the large $N$ saddle point eigenvalue distribution
\be
 \label{N^3/2:ansatz}
 u^{(a)}_j = \ii N^{1/2} t_j + v^{(a)}_j \, , \qquad \fm^{(a)}_j = \ii N^{1/2} \fn_j + \fp^{(a)}_j \, .
\ee
Observe that we have deformed the real integer fluxes $\fm_j$ into the complex plane in \eqref{N^3/2:ansatz}, anticipating a complex saddle point.
Moreover, the imaginary parts of $u_j^{(a)}$ and $\fm_j^{(a)}$ do \emph{not} depend on the index $a$.
At large $N$, we define the continuous functions
\bea
 t_j & \equiv t ( j / N ) \, , \qquad v_j^{(a)} \equiv v^{(a)} ( j / N ) \, , \\
 \fn_j & \equiv \fn ( j / N ) \, , \qquad \fp_j^{(a)} \equiv \fp^{(a)} ( j / N ) \, ,
\eea
and we introduce the normalized density of eigenvalues
\be
 \rho ( t ) = \frac{1}{N} \frac{\rd j}{\rd t} \, , \qquad \int \rd t \? \rho ( t ) = 1 \, .
\ee
We also define
\be
 \delta v ( t ) \equiv v_b ( t ) - v_a ( t ) \, , \qquad \delta \fp ( t ) \equiv \fp_b ( t ) - \fp_a ( t ) \, .
\ee
In taking the continuum limit the sums over $N$ become Riemann integrals, for example,
\be
 \sum_{j = 1}^N \to N \int \rd t \? \rho ( t ) \, .
\ee
Finally, we impose the constraint
\be
 \label{N^3/2:CS:const}
 k_{\text{CS}} \equiv \sum_{a = 1}^{|\cG|} k_a = 0 \, ,
\ee
as appropriate for quivers dual to M-theory on AdS$_4 \times  Y_7$ background, with $Y_7$ a Sasaki-Einstein five-manifold, and $N^{3/2}$ scaling.
We follow the logic of  \cite{Benini:2015eyy,Hosseini:2016tor}, to which we refer for more details about the method and assumptions.

\subsection{Chern-Simons}

Each group $a$ with CS level $k_a$ contributes to the index as
\bea
 \log Z_{\text{CS}} & = \ii k_a \sum_{i = 1}^{N} \fm_i u_i \\
 & \overset{N \gg 1}{=} - N^{3/2} k_a \int \rd t \? \rho ( t ) \left( \fn ( t  ) v_a ( t ) + t \? \fp_a ( t ) \right) \\
 & \qquad + \ii N k_a \int \rd t \? \rho ( t ) \fp_a ( t ) v_a ( t ) - \ii N^2 k_a \int \rd t \? \rho ( t ) t \? \fn ( t ) \, ,
\eea
where in the second equality we used the scaling ansatz \eqref{N^3/2:ansatz} and took the continuum limit.
Summing over nodes the last term vanishes because of \eqref{N^3/2:CS:const}.
Therefore, we obtain at large $N$
\be
 \label{CS:largeN:3/2}
 \log Z_{\text{CS}} = - N^{3/2} k_a \int \rd t \? \rho ( t ) \left( \fn ( t  ) v_a ( t ) + t \? \fp_a ( t ) \right) ,
\ee
reproducing \eqref{main:CS:largeN:3/2}.

\subsection{Chiral multiplet in bi-fundamental representation}

Let us consider first the contribution of a single chiral multiplet, in the $( \overline{\bf N}, {\bf N})$ representation of $\U(N)_a \times \U(N)_b$.
We denote the chemical potential and magnetic flux by $(\Delta, \fs)$.
 The polynomial piece in \eqref{logZ:chiral:ini} only participates in the cancellation of long-range forces or otherwise is subleading, as it can be checked explicitly using \eqref{long-range}. The relevant contribution from \eqref{logZ:chiral:ini} is
\be
 \label{logZ:chiral:dilog}
 \log Z_{(b , a)} = \sum_{i , j =1}^N \sum_{\ell = - \frac{|B_{i j}| - 1}{2}}^{\frac{|B_{i j}|-1}{2}} \sign (B_{i j})
 \Li_1 \left( e^{\ii (u_j^{(b)}- u_i^{(a)} + \Delta + \ell \epsilon )} \right) ,
\ee
where $B_{i j} = \fm_j^{(b)} - \fm_i^{(a)} - \fs + 1$.
We break $\sum_{i ,j} \to \sum_{i < j} + \sum_{i < j} + ( i \to j )$.

Observe that, in the large $N$ limit,
\be
 \label{logZ:chiral:i=j}
 \log Z_{(b , a)}^{i \to j}= N \int \rd t \? \rho ( t )
 \sum_{\ell = - \frac{|B ( t )| - 1}{2}}^{\frac{|B ( t )|-1}{2}} \sign \left( B ( t ) \right)
 \Li_1 \left( e^{\ii ( \delta v ( t ) + \Delta + \ell \epsilon )} \right) ,
\ee
where $B ( t ) \equiv \delta \fp ( t ) - \fs + 1$, is subleading.
\paragraph*{\emph{Digression.}}
Performing the saddle point approximation in the $(u-\fm)$-plane, we obtain
\be
Z ( \Delta, \fs | \epsilon ) \overset{N \gg 1}{\sim} \frac{Z_{(b,a)} ( u, \fm ; \Delta, \fs | \epsilon) }{\sqrt{\det \bH}} \bigg|_{\text{saddle point}} \, ,
\ee
where $\bH$ is the Hessian matrix
\be
 \bH =
  \begin{tikzpicture}[baseline=(current bounding box.center),
                      large/.style={font=\large}]
    \matrix (M)[matrix of math nodes, nodes in empty cells,
                left delimiter={(}, right delimiter={)},
                column sep={4.em,between origins},
                row sep={2.7em,between origins}
    ]{ \frac{\pd \log Z_{(b,a)}}{\pd u_i^{(a)} \pd u_j^{(a)}} & \frac{\pd \log Z_{(b,a)}}{\pd u_i^{(a)} \pd u_j^{(b)}}
    & \frac{\pd \log Z_{(b,a)}}{\pd u_i^{(a)} \pd \fm_j^{(a)}} & \frac{\pd \log Z_{(b,a)}}{\pd u_i^{(a)} \pd \fm_j^{(b)}} \\
    \frac{\pd \log Z_{(b,a)}}{\pd u_i^{(b)} \pd u_j^{(a)}} & \frac{\pd \log Z_{(b,a)}}{\pd u_i^{(b)} \pd u_j^{(b)}}
    & \frac{\pd \log Z_{(b,a)}}{\pd u_i^{(b)} \pd \fm_j^{(a)}} & \frac{\pd \log Z_{(b,a)}}{\pd u_i^{(b)} \pd \fm_j^{(b)}} \\
    \frac{\pd \log Z_{(b,a)}}{\pd \fm_i^{(a)} \pd u_j^{(a)}} & \frac{\pd \log Z_{(b,a)}}{\pd \fm_i^{(a)} \pd u_j^{(b)}}
    & \frac{\pd \log Z_{(b,a)}}{\pd \fm_i^{(a)} \pd \fm_j^{(a)}} & \frac{\pd \log Z_{(b,a)}}{\pd \fm_i^{(a)} \pd \fm_j^{(b)}} \\
    \frac{\pd \log Z_{(b,a)}}{\pd \fm_i^{(b)} \pd u_j^{(a)}} & \frac{\pd \log Z_{(b,a)}}{\pd \fm_i^{(b)} \pd u_j^{(b)}}
    & \frac{\pd \log Z_{(b,a)}}{\pd \fm_i^{(b)} \pd \fm_j^{(a)}} & \frac{\pd \log Z_{(b,a)}}{\pd \fm_i^{(b)} \pd \fm_j^{(b)}} \\
    };
    \draw[dashed](M-2-1.south west)--([xshift=2mm]M-2-4.south east);
    \draw[dashed]([xshift=1mm]M-1-3.north west)--([xshift=-0.9mm]M-4-2.south east);
  \end{tikzpicture}_{4N \times 4N} \, .
\ee
Let us set $\epsilon = 0$. Then, $Z_{(b,a)} (u , \fm ; \Delta, \fs | 0)$ becomes a linear function in the gauge magnetic fluxes $(\fm_i^{(a)}, \fm_{i}^{(b)} )$.
Explicitly, we can write
\be
 \log Z_{(b,a)} (u , \fm ; \Delta, \fs | 0) = \sum_{i , j =1}^N \big(  \fm_j^{(b)} - \fm_i^{(a)} - \fs + 1 \big) \Li_1 \left( e^{\ii (u_j^{(b)}- u_i^{(a)} + \Delta )} \right) ,
\ee
and, therefore,
\be
 \bH =
  \begin{tikzpicture}[baseline=(current bounding box.center),
                      large/.style={font=\large}]
    \matrix (M)[matrix of math nodes, nodes in empty cells,
                left delimiter={(}, right delimiter={)},
                column sep={4.em,between origins},
                row sep={2.7em,between origins}
    ]{ \frac{\pd \log Z_{(b,a)}}{\pd u_i^{(a)} \pd u_j^{(a)}} & \frac{\pd \log Z_{(b,a)}}{\pd u_i^{(a)} \pd u_j^{(b)}}
    & \bB \\
    \frac{\pd \log Z_{(b,a)}}{\pd u_i^{(b)} \pd u_j^{(a)}} & \frac{\pd \log Z_{(b,a)}}{\pd u_i^{(b)} \pd u_j^{(b)}} &
    \\
   \bB & & 0
    \\
    };
  \end{tikzpicture}_{4N \times 4N} \, .
\ee
where $\bB$ is the $2 N \times 2 N$ matrix appearing in the Jacobian of the Bethe approach \cite[(2.25)]{Benini:2015eyy}
\be
 \bB = \left(
  \begin{array}{cc}
   \frac{\pd \wt \cW_{(b,a)}}{\pd u_i^{(a)} \pd u_j^{(a)}} & \frac{\pd \wt \cW_{(b,a)}}{\pd u_i^{(a)} \pd u_j^{(b)}} \\
   \frac{\pd \wt \cW_{(b,a)}}{\pd u_i^{(b)} \pd u_j^{(a)}} & \frac{\pd \wt \cW_{(b,a)}}{\pd u_i^{(b)} \pd u_j^{(b)}} \\
  \end{array}
 \right)_{2 N \times 2 N} \, .
\ee
Here, we used the relations
\be\label{doubt}
 \frac{\pd \log Z_{(b,a)}}{\pd \fm_i^{(b)}} = \frac{\pd \wt \cW_{(b,a)}}{\pd u_i^{(b)}} \, , \qquad
 \frac{\pd \log Z_{(b,a)}}{\pd \fm_i^{(a)}} = \frac{\pd \wt \cW_{(b,a)}}{\pd u_i^{(a)}} \, .
\ee
Hence,
\be
 - \frac12 \log \det \bH
 = - \log \det \bB - \frac{\ii \pi}{2}
 = - \log \det_{i j} \partial^2_{u_i^{(a)} u_j^{(b)}} \wt \cW(u ; \Delta) - \frac{\ii \pi}{2} \, .
\ee
In the large $N$ limit, we find%
\footnote{The interested reader can find the details in \cite{Benini:2015eyy}.}
\be
 \label{cont:Jacobian}
 - \log \det \bB = - N \int \rd t \? \rho(t) \Li_1 ( e^{\ii ( \delta v ( t ) + \Delta )} ) \, .
\ee

Now, consider \eqref{logZ:chiral:i=j} in the $\epsilon \to 0$ limit. At large $N$, we obtain
\bea
 \label{cont:1:i=j}
 \log Z_{(b , a)}^{i \to j} & = \sum_{i = 1}^N \big( \fm_i^{(b)} - \fm_i^{(a)} - \fs + 1 \big) \Li_1 ( e^{\ii ( u_i^{(b)} - u_i^{(a)} + \Delta )} ) \\
 & = N \int \rd t \? \rho(t) (\delta \fp ( t) - \fs + 1) \Li_1 ( e^{\ii ( \delta v ( t ) + \Delta)} ) \, .
\eea
Putting together \eqref{cont:Jacobian} and \eqref{cont:1:i=j}, we then find the following contribution to the large $N$ twisted index
\be
 N (\delta \fp ( t) - \fs ) \int \rd t \? \rho(t) \Li_1 ( e^{\ii ( \delta v ( t ) + \Delta )} ) \, ,
\ee
that is subleading even in the \emph{tails}, where $(\delta v (t) , \delta \fp (t))$ are frozen to the constant boundary values $(- \Delta, \fs)$ up to exponentially small corrections. This is in contrast with the Bethe approach where the subtle contributions of the tails to the large $N$ twisted index must be included \cite{Benini:2015eyy}.

For finite $\epsilon$ there is no complication and we can easily ignore the contributions of $ \log Z_{(b , a)}^{i \to j}$ and $\log \det \bH$,
as they are suppressed by a power of $N^{-1/2}$ in comparison to the $N^{3/2}$ scaling of the index.

\paragraph*{\emph{Back to the refined case.}}
Let us now focus on
\bea
 \label{logZ:chiral:dilog:expand:i<j}
 \log Z_{(b , a)}^{i < j} & = \sum_{i<j}^N \sum_{\ell = - \frac{|B_{i j}| - 1}{2}}^{\frac{|B_{i j}|-1}{2}} \sign (B_{i j}) \sum_{n = 1}^{\infty} \frac{e^{\ii n ( u_j^{(b)} - u_i^{(a)} + \Delta + \ell \epsilon )}}{n} \\
 & = \sum_{n = 1}^\infty \frac{1}{n} \sum_{i<j}^N \frac{e^{\ii n B_{i j} \epsilon} - 1}{e^{\ii n \epsilon} - 1}
 e^{\ii n \left( u_j^{(b)} - u_i^{(a)} + \Delta - \frac{1}{2} ( B_{i j} - 1 ) \epsilon \right)} \, .
\eea
In the large $N$ limit we obtain
\be
 \label{Z:chi:i<j:N:3/2}
 \log Z_{(b , a)}^{i < j} = N^2 \sum_{n = 1}^{\infty} \frac{e^{\ii n ( \Delta + \frac{\epsilon}{2} \fs )}}{n \left( e^{\ii n \epsilon} - 1\right)} \int \rd t \? \rho ( t ) I_n ( t ) \, ,
\ee
where we defined
\bea
 I_n ( t ) & \equiv \int_t \rd t' \? \rho ( t' ) e^{- n N^{1/2} ( t' - t)}
 \left( e^{\ii n \epsilon \left( \ii N^{1/2} \left( \fn ( t' ) - \fn ( t ) \right) + \fp_b ( t') - \fp_a ( t ) - \fs + 1 \right)} - 1 \right) \\
 & \! \qquad \times e^{\ii n \left( v_b( t' ) - v_a ( t ) - \frac{\epsilon}{2} \left( \ii N^{1/2} \left( \fn ( t' ) - \fn ( t ) \right) + \fp_b ( t') - \fp_a ( t ) \right) \right)} \\
 & \equiv \int_t \rd t' \? \cI_n ( t', t ) \, .
\eea
We will assume that the functions $t\pm \frac{\epsilon}{2} \fn( t )$ are monotonic, so that the integral is convergent for  sufficiently large $N$. The integral can be evaluated by a saddle point approximation. Equivalently, performing integration by parts we obtain, up to sub-leading terms at large $N$,
\be
 \label{N^3/2:int:by:parts:1}
 \begin{aligned}
  I_n ( t ) & = - \frac{N^{-1/2}}{n} \cI_n ( t', t ) \Big|_t \\
  & + \frac{\ii \epsilon}{2} \int_t \rd t' \? \rho ( t' ) \cI_n ( t', t ) \fn'( t' ) \cot \left( \frac{n \epsilon}{2} \left( \ii N^{1/2} \left( \fn ( t' ) - \fn ( t ) \right) + \fp_b ( t' ) - \fp_a ( t ) - \fs + 1 \right) \right) .
 \end{aligned}
\ee
We again perform integration by parts on the second line of \eqref{N^3/2:int:by:parts:1} and find, at large $N$,
\bea
  I_n & = - \frac{N^{-1/2}}{n} \rho ( t' ) \cI_n ( t', t ) 
  \left[ 1 + \frac{\ii \epsilon}{2}  \fn' ( t' ) \cot \left( \frac{n \epsilon}{2} \left( \ii N^{1/2} \left( \fn ( t' ) - \fn ( t ) \right) + \fp_b ( t' ) - \fp_a ( t ) - \fs + 1 \right) \right) \right]_t \\
  & + \frac{\epsilon^2}{4} \int_t \rd t' \? \rho ( t' ) \cI_n ( t', t ) \fn' ( t' )^2 .
\eea
Via a repeated application of integration by parts we thus obtain the leading order result
\bea
 I_n ( t ) & = \frac{N^{-1/2}}{n} \rho( t )
 \left( e^{\ii n \epsilon ( \delta \fp ( t) + 1 - \fs )} - 1 \right)
 \left( e^{\ii n \left( \delta v ( t ) - \frac{\epsilon}{2} \delta \fp ( t )  \right)} \right) \\
 & \times \sum_{r = 0}^\infty \frac{(1 - \ii ) \left( \ii + (-1)^r  \right) }{2} \left( \frac{\epsilon \fn' ( t )}{2} \right)^r
 \cot^{\frac{1}{2} \left(1 - (-1)^r \right)} \left( \frac{n \epsilon}{2} \left( \delta \fp ( t ) + 1 - \fs \right) \right) .
\eea
Performing the sum $\sum_{r = 0}^\infty$ we find
\bea
 I_n ( t ) & = - \frac{N^{-1/2}}{n} \rho( t )
 \left( e^{\ii n \epsilon ( \delta \fp ( t) + 1 - \fs )} - 1 \right)
 \left( e^{\ii n \left( \delta v ( t ) - \frac{\epsilon}{2} \delta \fp ( t )  \right)} \right) \\
 & \times \left[ \frac{1}{\left( e^{- \ii n \epsilon \left( \delta \fp ( t ) + 1 - \fs \right)} - 1\right)} \frac{1}{\left( 1 + \frac{\epsilon}{2} \fn' ( t ) \right)}
 + \frac{1}{\left( e^{\ii n \epsilon \left( \delta \fp ( t ) + 1 - \fs \right)} - 1 \right)} \frac{1}{\left( 1 - \frac{\epsilon}{2} \fn' ( t ) \right)} \right] .
\eea
Therefore, in the large $N$ limit, \eqref{logZ:chiral:dilog} reduces to
\be
 \label{N^3/2:ini:chiral:final:i<j}
 \log Z_{(b , a)}^{i<j} = N^{3/2} \sum_{n = 1}^{\infty}
 \frac{e^{\ii n \left( \delta v ( t ) + \Delta + \frac{\epsilon}{2} \right)}}{n^2 \left( e^{\ii n \epsilon} - 1 \right)}
  \int \rd t \? \rho( t )^2
 \left(
 \frac{e^{\frac{\ii n \epsilon}{2} \left( \delta \fp ( t ) + 1 - \fs \right)}}{1 + \frac{\epsilon }{2} \fn'(t)}
 - \frac{e^{- \frac{\ii n \epsilon}{2} \left( \delta \fp ( t ) + 1 - \fs \right)}}{1 - \frac{\epsilon }{2} \fn'(t)}
 \right) .
\ee
Now, it remains to perform the sum over $n$. We obtain%
\footnote{Here, we used the relation $\Li_2 ( e^{\ii u} ) = \sum_{m = 1}^\infty \frac{e^{\ii m u}}{m^2}$ and exchanged the order of summations.}
\bea
 \label{N^3/2:chiral:final:i<j}
 \log Z_{(b , a)}^{i<j}= - N^{3/2}
 \sum_{n = 1}^\infty
  \int \rd t \? \rho( t )^2
 \Bigg( &
 \frac{\Li_2 \left( e^{\ii \left( \delta v ( t ) + \Delta + \frac{\epsilon}{2} ( \delta \fp ( t ) - \fs + 2 n \right)} \right)}{1 + \frac{\epsilon }{2} \fn'(t)} \\
 & - \frac{\Li_2 \left( e^{\ii \left( \delta v ( t ) + \Delta - \frac{\epsilon}{2} \left( \delta \fp ( t ) - \fs - 2 ( n - 1 ) \right) \right)} \right)}{1 - \frac{\epsilon }{2} \fn'(t)}
 \Bigg) .
\eea
The summation $\sum_{i > j}$ in \eqref{logZ:chiral:dilog} is similar to \eqref{N^3/2:chiral:final:i<j} and it reads%
\footnote{To have a convergent integral we need first to use the inversion formul\ae\;\eqref{Li:inversion:g}.
The polynomial pieces only enter in the cancellation of long-range forces. See 
\cite{Benini:2015eyy,Hosseini:2016tor} for more details.}
\be
 \label{N^3/2:chiral:final:i>j}
 \begin{aligned}
 \log Z_{(b , a)}^{i>j} = N^{3/2}
 \sum_{n = 1}^\infty
  \int \rd t \? \rho( t )^2
 \Bigg( &
 \frac{\Li_2 \left( e^{- \ii \left( \delta v ( t ) + \Delta + \frac{\epsilon}{2} \left( \delta \fp ( t ) - \fs - 2 ( n - 1 ) \right) \right)} \right)}{1 + \frac{\epsilon }{2} \fn'(t)} \\
 & - \frac{\Li_2 \left( e^{- \ii \left( \delta v ( t ) + \Delta - \frac{\epsilon}{2} \left( \delta \fp ( t ) - \fs + 2 n \right) \right)} \right)}{1 - \frac{\epsilon }{2} \fn'(t)}
 \Bigg) \, .
 \end{aligned}
\ee
Putting together \eqref{N^3/2:chiral:final:i<j} and \eqref{N^3/2:chiral:final:i>j}, we finally arrive at the following compact expression for the contribution of a chiral multiplet to the \emph{refined} twisted index at large $N$
\bea
 \label{N^3/2:chiral:FINAL}
 \log Z_{(b , a)} = N^{3/2}
  \int \rd t \? \rho( t )^2
 \bigg( &
 \frac{\psi \left( - \delta v ( t ) - \Delta - \frac{\epsilon}{2} \left( \delta \fp ( t ) - \fs \right) ; \epsilon \right)}{1 + \frac{\epsilon }{2} \fn'(t)} \\
 & - \frac{\psi \left( - \delta v ( t ) - \Delta + \frac{\epsilon}{2} \left( \delta \fp ( t ) - \fs + 2 \right) ; \epsilon \right)}{1 - \frac{\epsilon }{2} \fn'(t)}
 \bigg) \, .
\eea
Here, we introduced the ``elliptic dilogarithm'' function
\be
 \label{def:psi:dilog}
 \psi ( \Delta ; \epsilon) \equiv \sum_{n = 0}^\infty \left( \Li_2 \left(e^{\ii ( \Delta + n \epsilon )}\right) - \Li_2 \left(e^{- \ii \left( \Delta - ( n +1 ) \epsilon \right)}\right) \right) .
\ee

\paragraph*{Asymptotic expansion around $\epsilon = 0$ --}
Starting with \eqref{N^3/2:chiral:FINAL} we can write down the following asymptotic expansion around $\epsilon = 0$
\bea
 \label{N^3/2:chiral:asymp:ini}
 \log Z_\chi = N^{3/2} \int \rd t \? \rho ( t )^2
 \Bigg[
 & - \sum_{l = 0}^{3} \frac{1}{\pi^l}
 g_l \left( - \pi \left( \delta \fp ( t ) - \fs \right) \right)
 g_{3 - l} \left( \delta v ( t ) + \Delta \right)
 \sum_{n = 0}^{\infty} \left(\frac{\epsilon \? \fn' ( t )}{2} \right)^{2 n} \\
 & + \frac{1}{\pi^3} g_3 \left( - \pi \left( \delta \fp ( t ) - \fs \right) \right) \fn' ( t )^{-2} \\
 & + \frac{1}{\pi^2} g_2 \left( - \pi \left( \delta \fp ( t ) - \fs \right) \right) g_1 \left( \delta v ( t ) + \Delta \right) \fn' ( t )^{-1} \Bigg] \, ,
 \eea
where we used the inversion formul\ae
\be
 \label{Li:inversion:g}
 \Li_s (e^{\ii u} ) + ( - 1)^s \Li_s (e^{-\ii u} ) = - \frac{(2 \pi \ii)^s}{s!} B_s \left( \frac{u}{2 \pi} \right) \equiv \ii^{s -2} g_s ( u ) \, ,
\ee
for $0 < \re ( u ) < 2 \pi$.
Here, the polynomial functions $g_s (u)$ are related to the Bernoulli polynomials $B_s (u)$ via \eqref{Li:inversion:g} and for $s=1,2,3$ are explicitly given in \eqref{main:g123}. 
Using $\sum_{n = 0}^\infty x^{2 n} = (1 - x^2)^{-1}$, \eqref{N^3/2:chiral:asymp:ini} is simplified to
\be
\label{N^3/2:cardy:FINAL:chi}
\log Z_{(b , a)} = N^{3/2} \int \rd t \? \rho ( t )^2
 \left(
 \frac{\cG ( \delta v ( t ) + \Delta, - \delta \fp ( t ) + \fs , \epsilon )}{1 + \frac{\epsilon}{2} \fn' ( t )}
 - \frac{\cG ( \delta v ( t ) + \Delta, \delta \fp ( t ) - \fs + 2, \epsilon )}{1 - \frac {\epsilon} {2} \fn' ( t )}
 \right) ,
\ee
where we defined
\be
 \label{def:cG:poly}
 \cG (u, \fs, \epsilon) \equiv \frac{1}{\epsilon} \left[
 g_3 ( u )
 - \left( \frac{\epsilon}{2 \pi}\right) g_2 ( u ) g_1( \pi \fs)
 + \left(\frac{\epsilon}{2 \pi}\right)^2 g_1 ( u ) g_2 ( \pi \fs)
 - \left(\frac{\epsilon}{2 \pi}\right)^3 g_3( \pi \fs)
 \right] .
\ee
Let us define the \emph{equivariant} quantities
\bea
 & w^{(1)} ( t ) \equiv \ii t + \frac{\ii \epsilon}{2} \fn ( t ) \, , && \tv_{a}^{(1)} ( t ) \equiv \frac{1}{\omega} \left( v_a + \frac{\epsilon}{2} \fp_a ( t ) \right) , \\
 & \bbDelta_I^{(1)} \equiv \frac{1}{\omega} \left( \Delta_I + \pi ( \omega - 1 ) + \frac{\epsilon}{2} ( 1 - \fs_I ) \right) , \qquad && \epsilon^{(1)} \equiv \epsilon \, , \\
 & w^{(2)} ( t ) \equiv \ii t - \frac{\ii \epsilon}{2} \fn ( t ) \, , && \tv_{a}^{(2)} ( t ) \equiv \frac{1}{\omega} \left( v_a - \frac{\epsilon}{2} \fp_a ( t ) \right) , \\
 & \bbDelta_I^{(2)} \equiv \frac{1}{\omega} \left( \Delta_I + \pi ( \omega - 1 ) - \frac{\epsilon}{2} ( 1 - \fs_I ) \right) , \qquad && \epsilon^{(2)} \equiv - \epsilon \, ,
\eea
with
\be
 \label{def:omega:epsilon}
 \omega \equiv \sqrt{1 + \left( \frac{\epsilon}{2 \pi} \right)^2} \, .
\ee
Remarkably, \eqref{def:cG:poly} obeys the following relations
\bea
 \cG \left( \delta v ( t ) + \Delta , \fs , \epsilon \right) & = \frac{\omega^3}{\epsilon} g_3 \left( \delta \tv^{(1)} ( t ) + \bbDelta^{(1)} \right) \, , \\
 \cG \left( \delta v ( t ) + \Delta , 2 - \fs , \epsilon \right) & = \frac{\omega^3}{\epsilon} g_3 \left( \delta \tv^{(2)} ( t ) + \bbDelta^{(2)} \right)  \, .
\eea
Then, \eqref{N^3/2:cardy:FINAL:chi} takes the following factorized form
\be
 \label{N^3/2:cardy:factorized:chi}
 \log Z_{(b , a)} = \ii \omega^3 N^{3/2} \sum_{\sigma = 1}^2 \frac{1}{\epsilon^{(\sigma)}} \int \rd t \? \rho ( t )^2
 \frac{g_3 ( \delta \tv^{(\sigma)} ( t ) + \bbDelta^{(\sigma)} )}{w'^{(\sigma)}( t )} \, .
\ee

Thus, a pair of bi-fundamental chiral multiplets with chemical potentials and fluxes $(\Delta_{I}, \fs_I)$, $I = (a,b), (b,a)$,
transforming, respectively, in the $({\bf N},\overline{\bf N})$ and $(\overline{\bf N},{\bf N})$ representation of $\U(N)_a \times \U(N)_b$ contributes to the large $N$ refined index
\be
 \label{N^3/2:cardy:factorized:bifund}
 \log Z_{(a \leftrightarrow b)} = \ii \omega^3 N^{3/2}
  \sum_{\substack{I = (b , a): + \\ I = (a , b): -}}
  \sum_{\sigma =1}^2
  \frac{1}{\epsilon^{(\sigma)}}
  \int \rd t \? \rho ( t )^2 \?
  \frac{g_3 ( \pm \delta \tv^{(\sigma)} ( t ) + \bbDelta_{I}^{(\sigma)} )}{w'^{(\sigma)}( t )} \, ,
\ee
reproducing \eqref{main:N^3/2:cardy:factorized:bifund}.

The analysis for the large $N$ contribution of a chiral multiplet in the adjoint representation of $\U(N)_a$ is very similar.
Setting
\be
 \delta \tv^{(\sigma)} ( t ) = 0 \, , \qquad \bbDelta^{(\sigma)} \equiv \bbDelta_{(a,a)}^{(\sigma)} \, ,
\ee
in \eqref{N^3/2:cardy:factorized:chi}, it reads
\be
 \label{N^3/2:cardy:factorized:adj}
 \log Z_{(a , a)} = \ii \omega^3 N^{3/2}
 \sum_{\sigma =1}^2
 \frac{g_3 ( \bbDelta_{(a,a)}^{(\sigma)} )}{\epsilon^{(\sigma)}}
 \int \rd t \? \frac{\rho ( t )^2}{w'^{(\sigma)}( t )} \, ,
\ee
reproducing \eqref{main:N^3/2:cardy:factorized:adj}.

\subsection{Chiral multiplet in (anti-)fundamental representation}

A chiral multiplet transforming in the fundamental representation of $\U(N)_a$ contributes to the refined twisted index as%
\footnote{Here, we applied the inversion formula \eqref{Li:inversion:g} to \eqref{logZ:chiral:ini}.}
\be
 \label{logZ:fund:tot}
 \log Z_{\text{fund}} = \sum_{i = 1}^N \sum_{\ell = - \frac{|B_{i}| - 1}{2}}^{\frac{|B_{i}|-1}{2}} \sign (B_{i})
 \left( \Li_1 \left( e^{\ii ( - u_i^{(a)} - \Delta - \ell \epsilon )} \right) - \frac{\ii}2 g_1 ( u_i^{(a)} + \Delta + \ell \epsilon ) \right) .
\ee
Here, $B_i \equiv \fm_i - \fs + 1$ and $( \Delta, \fs)$ label the chemical potential and the magnetic flux, respectively.
Consider first the term
\be
  - \frac{\ii}2 \sum_{i = 1}^N \sum_{\ell = - \frac{|B_{i}| - 1}{2}}^{\frac{|B_{i}|-1}{2}} \sign (B_{i}) g_1 ( u_i^{(a)} + \Delta + \ell \epsilon )
  = - \frac{\ii}{2} \sum_{i = 1}^N (\fm_i - \fs + 1 ) ( u_i^{(a)} + \Delta - \pi ) .
\ee
Substituting the ansatz \eqref{N^3/2:ansatz} and taking the continuum limit, we can write
\be
 \label{Z:fund:g1:large:N}
 \begin{aligned}
 & + \frac{\ii}{2} N^2 \int \rd t \? \rho ( t ) \? t \? \fn ( t )
 + \frac{1}{2} N^{3/2} \int \rd t \? \rho ( t ) \left(  t ( \fp_a ( t ) - \fs + 1 ) - \fn ( t ) ( \pi - \Delta - v_a ( t ) ) \right)  \\
 & + \frac{\ii}{2} N \int \rd t \? \rho ( t ) ( \fp_a ( t ) - \fs + 1 ) ( \pi - \Delta - v_a ( t ) ) \, .
 \end{aligned}
\ee
Let us now focus on
\bea
 \label{logZ:fund:dilog}
 \sum_{i = 1}^N \sum_{\ell = - \frac{|B_{i}| - 1}{2}}^{\frac{|B_{i}|-1}{2}} \sign (B_{i}) \Li_1 \left( e^{\ii ( - u_i^{(a)} - \Delta - \ell \epsilon )} \right) .
\eea
We need to extrapolate this formula to complex large values of $B_i$.
The summation variable $\ell$ will also describe a path in the complex plane interpolating between points that, at large $N$, scale as $\pm \frac{\ii}{2} \sqrt{N} \? \fn(t)$.
The argument of $\Li_1$ then interpolates between points that scale as $e^{ \sqrt{N} ( t \pm \frac{\epsilon}{2} \fn (t) )}$.
To evaluate $\sum_{\ell = - \frac{|B_{i}| - 1}{2}}^{\frac{|B_{i}|-1}{2}}$, we consider  distinct regions that are determined by $\sign ( t \pm \frac{\epsilon}{2} \fn (t) )$.
Explicitly, we consider
\bea
 \text{(i)} & \qquad t - \frac{\epsilon}{2} \fn ( t ) < 0 \, , \quad &&  t + \frac{\epsilon}{2} \fn ( t ) < 0 \, , \\
 \text{(ii)} & \qquad t - \frac{\epsilon}{2} \fn ( t ) > 0 \, , \quad &&  t + \frac{\epsilon}{2} \fn ( t ) < 0 \, , \\
 \text{(iii)} & \qquad t - \frac{\epsilon}{2} \fn ( t ) > 0 \, , \quad &&  t + \frac{\epsilon}{2} \fn ( t ) > 0 \, ,
\eea
and a fourth region that can be similarly studied.
In region (i), the contribution of \eqref{logZ:fund:dilog} to the large $N$ refined twisted index is subleading,
since the argument of $\Li_1$ is exponentially suppressed for all values of $\ell$.
Consider now region (iii). At large $N$, the argument of $\Li_1$ blows up for all values of $\ell$ and we obtain
\bea
 \label{Z:fund:Li1:large:N:iii}
 & - \ii N^2 \int_{\text{(iii)}} \rd t \? \rho ( t ) \? t \? \fn ( t )
 - N^{3/2} \int_{\text{(iii)}} \rd t \? \rho ( t ) \left( t ( \fp_a ( t ) - \fs + 1 ) - \fn ( t ) ( \pi - \Delta - v_a ( t ) ) \right)
 + \cO ( N ) \, .
\eea
In region (ii), we restrict the sum over $\ell$ to the values where the argument of $\Li_1$ blows up,
namely to the values $\ell \lessgtr - \frac{1}{\epsilon} ( u_i + \Delta + 2 \pi c )$ for some $c$ of order one, for $- \fn( t ) \lessgtr 0$.
Thus, summing over $\ell$, the large $N$ limit of \eqref{logZ:fund:dilog} in region (ii) reads
\bea
 \label{Z:fund:Li1:large:N:ii}
 & - \frac{\ii}{2} N^2 \int_{\text{(ii)}} \rd t \? \rho ( t ) \? t \? \fn ( t )
 + \frac{\ii}{2 \epsilon} N^2 \int_{\text{(ii)}} \rd t \? \rho ( t ) \left( t^2 + \frac{\epsilon^2}{4} \fn ( t )^2 \right) \\
 & - \frac{1}{2} N^{3/2} \int_{\text{(ii)}} \rd t \? \rho ( t ) \left( t ( \fp_a ( t ) - \fs + 1 ) - \fn ( t ) ( \pi - \Delta - v_a ( t) ) \right) \\
 & - \frac{1}{\epsilon} N^{3/2} \int_{\text{(ii)}} \rd t \? \rho ( t ) \left( t ( \pi - \Delta - v_a ( t ) ) - \frac{\epsilon^2}{4} \fn ( t ) ( \fp_a ( t ) - \fs + 1 ) \right) + \cO ( N ) \, .
\eea
Putting together \eqref{Z:fund:g1:large:N}, \eqref{Z:fund:Li1:large:N:iii}, and \eqref{Z:fund:Li1:large:N:ii}, we find
\bea
 \label{Z:fund:largeN}
  \log Z_{\text{fund}} & =
  \frac{\ii}{2} N^2 \int_{\text{(i)}} \rd t \? \rho ( t ) \? t \? \fn ( t )
  + \frac{\ii}{2 \epsilon} N^2 \int_{\text{(ii)}} \rd t \? \rho ( t ) \left( t^2 + \frac{\epsilon^2}{4} \fn ( t )^2 \right)
  - \frac{\ii}{2} N^2 \int_{\text{(iii)}} \rd t \? \rho ( t )  \? t \? \fn ( t ) \\
  & + \frac12 N^{3/2} \int_{\text{(i)}} \rd t \? \rho ( t ) \left( t ( \fp_a ( t ) - \fs +1 ) - \fn ( t ) ( \pi - \Delta - v_a ( t ) ) \right) \\
  & - \frac{1}{\epsilon} N^{3/2} \int_{\text{(ii)}} \rd t \? \rho ( t ) \left( t ( \pi - \Delta - v_a ( t ) ) - \frac{\epsilon^2}{4} \fn ( t )  ( \fp_a ( t ) - \fs + 1 )  \right) \\
  & - \frac12 N^{3/2} \int_{\text{(iii)}} \rd t \? \rho ( t )  \left( t ( \fp_a ( t ) - \fs + 1 ) - \fn ( t ) ( \pi - \Delta - v_a ( t ) ) \right) + \cO ( N ) \, .
\eea
The contribution of a chiral field with chemical potential $\wt \Delta$ and magnetic flux $\tilde \fs$, transforming in the anti-fundamental representation of $\U(N)_a$ is similarly given by
\bea
 \label{Z:anti-fund:largeN}
 \log Z_{\text{anti-fund}} & =
 - \frac{\ii}{2} N^2 \int_{\text{(i)}} \rd t \? \rho ( t ) \? t \? \fn ( t )
 - \frac{\ii}{2 \epsilon} N^2 \int_{\text{(ii)}} \rd t \? \rho ( t ) \left( t^2 + \frac{\epsilon^2}{4} \fn ( t )^2 \right)
 + \frac{\ii}{2} N^2 \int_{\text{(iii)}} \rd t \? \rho ( t )  \? t \? \fn ( t ) \\
 & - \frac12 N^{3/2} \int_{\text{(i)}} \rd t \? \rho ( t ) \left( t ( \fp_a ( t ) + \tilde \fs -1 ) + \fn ( t ) ( \pi - \wt \Delta + v_a ( t ) ) \right) \\
 & - \frac{1}{\epsilon} N^{3/2} \int_{\text{(ii)}} \rd t \? \rho ( t ) \left( t ( \pi - \wt \Delta + v_a ( t ) ) + \frac{\epsilon^2}{4} \fn ( t )  ( \fp_a ( t ) + \tilde \fs - 1 )  \right) \\
 & + \frac12 N^{3/2} \int_{\text{(iii)}} \rd t \? \rho ( t )  \left( t ( \fp_a ( t ) + \tilde \fs - 1 ) + \fn ( t ) ( \pi - \wt \Delta + v_a ( t ) ) \right) + \cO ( N ) \, .
\eea
The first lines of \eqref{Z:fund:largeN} and \eqref{Z:anti-fund:largeN} cancel each other out for theories with total number of fundamentals equal to total number of anti-fundamentals, which we need to assume for consistency.
Therefore, keeping only $\cO(N^{3/2} )$ terms in \eqref{Z:fund:largeN} and \eqref{Z:anti-fund:largeN}, we find
\bea
 \log Z_{\text{fund}} & = - \frac{1}{2} N^{3/2} \sum_{\sigma = 1}^2 \int \rd t \? \rho( t )  \left( ( \pi - \Delta - v_a ( t ) ) - \frac{\epsilon^{(\sigma)}}{2} ( 1 - \fs + \fp_a ( t ) ) \right) \frac{| w^{(\sigma)} |}{\epsilon^{(\sigma)}} \, , \\
 \log Z_{\text{anti-fund}} & = - \frac{1}{2} N^{3/2} \sum_{\sigma = 1}^2 \int \rd t \? \rho( t ) \left( ( \pi - \wt \Delta + v_a ( t ) ) - \frac{\epsilon^{(\sigma)}}{2} ( 1 - \wt \fs - \fp_a ( t ) ) \right) \frac{ | w^{(\sigma)} | }{\epsilon^{(\sigma)}} \, ,
\eea
where we used the $A$-gluing parameterization
\bea
 w^{(1)} ( t ) & = \ii t + \ii \frac{\epsilon}{2} \fn ( t ) \, , \qquad \epsilon^{(1)} = \epsilon \, , \\
 w^{(2)} ( t ) & = \ii t - \ii \frac{\epsilon}{2} \fn ( t ) \, , \qquad \epsilon^{(2)} = - \epsilon \, .
\eea
This reproduces \eqref{main:N^3/2:fund} and \eqref{main:N^3/2:anti-fund}, after relabeling $( \Delta, \fs ) \to ( \Delta_a, \fs_a)$ and $( \wt \Delta, \tilde \fs ) \to ( \wt \Delta_a, \tilde \fs_a)$.
 
\subsection{Vector multiplet}

The contribution of a vector multiplet can be obtained simply via 
\be
 \label{chi:map:vec}
 \log Z_\cV = \log Z_\chi \Big|_{\Delta_{(a,a)} = 2 \pi, \, \fs_{(a,a)} = 2} \, .
\ee
From \eqref{N^3/2:cardy:factorized:adj} we obtain the following expression for a vector multiplet 
\be
\label{N^3/2:cardy:FINAL:vec}
\log Z_{\cV} = \frac{\pi}{12} N^{3/2} \int \rd t \? \rho ( t )^2
 \left(
 \frac{\epsilon - 2 \pi}{1 + \frac{\epsilon}{2} \fn' ( t )}
 - \frac{\epsilon + 2 \pi}{1 - \frac {\epsilon} {2} \fn' ( t )}
 \right) .
\ee

\section{Saddle point of the ADHM refined twisted index}
\label{app:ADHM}

This appendix contains the large $N$ saddle point for the ADHM quiver described in section \ref{sect:ADHM}.
The eigenvalue density distribution $\rho(t)$ and the magnetic flux $\fn( t )$, which extremize the refined index \eqref{ADHM:Z:factor},
are piece-wise functions supported on $[t_\ll , t_\gg]$.
For ease of notation, let us define
\bea
 G_3 ( v ) \equiv v_1 v_2 v_3 \, , \qquad
 \delta_+^{(\sigma)} \equiv \frac{1}{r} \left( \Delta_3^{(\sigma)} + \frac{r}{2} \Delta_m \right) \, , \qquad
 \delta_-^{(\sigma)} \equiv \frac{1}{r} \left( \Delta_3^{(\sigma)} - \frac{r}{2} \Delta_m \right) .
\eea
We define the inner interval as
\be
t_< ~ \text{ s.t. } ~ t_< - \frac{\epsilon}{2} \fn ( t_< ) = 0 \;,\qquad\qquad t_> ~ \text{ s.t. } ~ t_> + \frac{\epsilon}{2} \fn ( t_> ) = 0 \, .
\ee
Schematically, we have:
\begin{center}
\begin{tikzpicture}[scale=2]
\draw (-2.,0) -- (2.,0);
\draw (-2,-.05) -- (-2, .05); \draw (-0.7,-.05) -- (-0.7, .05); \draw (0.7,-.05) -- (0.7, .05); \draw (2,-.05) -- (2, .05);
\node [below] at (-2,0) {$t_\ll$}; \node [below] at (-2,-.3) {$\rho=0$};
\node [below] at (-0.7,0) {$t_<$}; \node [below] at (-0.7,-.3) {$t - \frac{\epsilon}{2} \fn = 0$};
\node [below] at (0.7,0) {$t_>$}; \node [below] at (0.7,-.3) {$ t + \frac{\epsilon}{2} \fn = 0$};
\node [below] at (2,0) {$t_\gg$}; \node [below] at (2,-.3) {$\rho=0$};
\end{tikzpicture}
\end{center}
The transition points are at
\bea
 t_{\ll} & = - \sum_{\sigma = 1}^2 \sqrt{\frac{ \Delta_1^{(\sigma)} \Delta_2^{(\sigma)} \delta_+^{(\sigma)} }
 {2 r \? \delta_-^{(\sigma)} }} \, , \qquad &&
 t_{<} = - \epsilon \? \sqrt{ G_3 ( \Delta^{(1)}) } \sum_{\sigma = 1}^2 \frac{1}{\epsilon^{(\sigma)}} \sqrt{\frac{ \delta_+^{(\sigma)} }{2 r \? \Delta_3^{(\sigma)}  \delta_-^{(\sigma)} }} \, , \\
 t_{>} & = - \epsilon \? \sqrt{ G_3 ( \Delta^{(2)}) } \sum_{\sigma = 1}^2 \frac{1}{\epsilon^{(\sigma)}} \sqrt{\frac{ \delta_-^{(\sigma)} }{2 r \? \Delta_3^{(\sigma)}  \delta_+^{(\sigma)} }} \, , &&
 t_{\gg} = \sum_{\sigma = 1}^2 \sqrt{\frac{ \Delta_1^{(\sigma)} \Delta_2^{(\sigma)} \delta_-^{(\sigma)} }
 {2 r \? \delta_+^{(\sigma)} }} \, .
\eea
In the left segment $ t \mp \frac{\epsilon}{2} \fn ( t ) < 0$, and we have
\bea
 \label{LS:rho:n}
 \rho ( t ) & = 4 r \left( \sum_{\sigma = 1}^2 \sqrt{ \frac{G_3 ( \Delta^{(\sigma ) } ) }{\delta_-^{(\sigma)}} } \right)^{-2}
 \Bigg( t + \frac{1}{2} \sum_{\sigma = 1}^2 \sqrt{ \frac{2 \? \delta_+^{(\sigma)} \Delta_1^{(\sigma)} \Delta_2^{(\sigma)} }
 {r \? \delta_-^{(\sigma)} } } \? \Bigg) , \\
 \fn ( t ) & = 2 \left( \sum_{\sigma = 1}^2 \sqrt{ \frac{G_3 ( \Delta^{(\sigma ) } ) }{\delta_-^{(\sigma)}} } \right)^{-1} \sum_{\sigma = 1}^2 \frac{1}{\epsilon^{(\sigma)}} \sqrt{ \frac{G_3 ( \Delta^{(\sigma)} )}{\delta_-^{(\sigma)}} } \? t \\
 & - 2 \left( \sum_{\sigma = 1}^2 \sqrt{ \frac{G_3 ( \Delta^{(\sigma ) } ) }{\delta_-^{(\sigma)}} } \right)^{-1} \prod_{\sigma = 1}^2 \sqrt{ \frac{G_{3} ( \Delta^{(\sigma)} )}{\delta_-^{(\sigma)}}}
 \sum_{\sigma = 1}^2 \frac{1}{\epsilon^{(\sigma)}} \sqrt{ \frac{2 \? \delta_+^{(\sigma)}}{r \? \Delta_3^{(\sigma)}} } \, .
\eea
In the middle segment $ t - \frac{\epsilon}{2} \fn ( t ) > 0$, $ t + \frac{\epsilon}{2} \fn ( t ) < 0$, and we find
\bea
 \rho ( t ) & = r \left( \frac{G_3 ( \Delta^{(1)})} {1 + \frac{ \epsilon }{2} \fn' ( t )} - \frac{G_3 ( \Delta^{(2)}) }{ 1 - \frac{\epsilon}{2} \fn ' ( t ) } \right)
 \left( \delta_+^{(2)} \left( t - \frac{\epsilon}{2} \fn ( t ) \right) + \delta_-^{(2)} \left( t + \frac{\epsilon}{2} \fn ( t ) \right) + \mu \epsilon \right) , \\
 \fn ( t ) & = \frac{2}{\epsilon} \left( \frac{G_3 ( \Delta^{(1)} )}{\delta_-^{(1)}} + \frac{G_3 ( \Delta^{(2)} )}{\delta_+^{(2)}} \right)^{-1}\left( c_1 t + c_2 + \sqrt{ c_3 t^2 + c_4 t + c_5} \right) ,
\eea
with
\bea
 c_1 & = \frac{G_3 ( \Delta^{(1)} )}{\delta_-^{(1)}} - \frac{G_3 ( \Delta^{(2)} )}{\delta_+^{(2)}} \, , \qquad
 c_2 = - \frac{G_3 ( \Delta^{(1)} ) G_3 ( \Delta^{(2)} )}{\delta_-^{(1)} \delta_+^{(2)}}
 \sum_{\sigma = 1}^2 \sqrt{ \frac{2 \? \delta_+^{(\sigma)} \delta_-^{(\sigma)} }{r \? \Delta_3^{(\sigma)} G_3 (\Delta^{(\sigma)})} }\, , \\
 c_3 & = - 4 \frac{G_3 ( \Delta^{(1)} ) G_3 ( \Delta^{(2)} )}{\delta_-^{(1)} \delta_+^{(2)}} \, , \\
 c_4 & = - 4 \frac{ G_3 ( \Delta^{(1)} ) G_3 ( \Delta^{(2)} ) }{ \delta_-^{(1)} \delta_+^{(2)} }
 \Bigg( \sqrt{ \frac{ 2 \? \delta_+^{(1)} \Delta_1^{(1)} \Delta_2^{(1)} }{ r \? \delta_-^{(1)} } }
 - \sqrt{ \frac{ 2 \? \delta_-^{(2)} \Delta_1^{(2)} \Delta_2^{(2)} }{ r \? \delta_+^{(1)} } } \Bigg) \, , \\
 c_5 & = \frac{2}{r} \frac{ G_3 ( \Delta^{(1)} ) G_3 ( \Delta^{(2)} ) }{ \delta_-^{(1)} \delta_+^{(2)} } \\
 & \times \left[
 \Bigg( \sqrt{ \frac{ \delta_+^{(1)} \Delta_1^{(1)} \Delta_2^{(1)} }{ r \? \delta_-^{(1)} } }
 + \sqrt{ \frac{ \delta_-^{(2)} \Delta_1^{(2)} \Delta_2^{(2)} }{ r \? \delta_+^{(1)} } } \Bigg)^2
 - \Bigg( \frac{(\delta_+^{(1)} - \delta_-^{(1)} ) \Delta_1^{(1)} \Delta_2^{(1)} }{ \delta_-^{(1)} }
 - \frac{ ( \delta_+^{(2)} - \delta_-^{(2)} ) \Delta_1^{(2)} \Delta_2^{(2)} }{ \delta_+^{(2)} } \Bigg) \right] ,
\eea
and
\be
 \mu = \sum_{\sigma = 1}^2 \frac{1}{\epsilon^{(\sigma)}} \sqrt{ 2 r \? \delta_-^{(\sigma)} \delta_+^{(\sigma)} \Delta_1^{(\sigma)} \Delta_2^{(\sigma)}} \, .
\ee
In the right segment $t \mp \frac{\epsilon}{2} \fn ( t ) > 0$, and we have
\bea
 \label{RS:rho:n}
 \rho ( t ) & = - 4 r \left( \sum_{\sigma = 1}^2 \sqrt{ \frac{G_3 ( \Delta^{(\sigma ) } ) }{\delta_+^{(\sigma)}} } \right)^{-2}
 \Bigg( t - \frac{1}{2} \sum_{\sigma = 1}^2 \sqrt{ \frac{2 \? \delta_-^{(\sigma)} \Delta_1^{(\sigma)} \Delta_2^{(\sigma)} }
 {r \? \delta_+^{(\sigma)} } } \? \Bigg) \, , \\
 \fn ( t ) & = 2 \left( \sum_{\sigma = 1}^2 \sqrt{ \frac{G_3 ( \Delta^{(\sigma ) } ) }{\delta_+^{(\sigma)}} } \right)^{-1} \sum_{\sigma = 1}^2 \frac{1}{\epsilon^{(\sigma)}} \sqrt{ \frac{G_3 ( \Delta^{(\sigma)} )}{\delta_+^{(\sigma)}} } \? t \\
 & + 2 \left( \sum_{\sigma = 1}^2 \sqrt{ \frac{G_3 ( \Delta^{(\sigma ) } ) }{\delta_+^{(\sigma)}} } \right)^{-1} \prod_{\sigma = 1}^2 \sqrt{ \frac{G_{3} ( \Delta^{(\sigma)} )}{\delta_+^{(\sigma)}}}
 \sum_{\sigma = 1}^2 \frac{1}{\epsilon^{(\sigma)}} \sqrt{ \frac{2 \? \delta_-^{(\sigma)}}{r \? \Delta_3^{(\sigma)}} } \, .
\eea

\section{ABJM unrefined twisted index at large $N$}
\label{app:ABJM}

In this appendix we provide the detailed form of the large $N$ saddle point for the ABJM \emph{unrefined} twisted index.
For notational convenience, we define
\bea
 G_1 ( v ) & \equiv \sum_{I = 1}^4 v_I \, , \qquad
 G_2 ( v ) \equiv v_1 v_2 - v_3 v_4 \, , \qquad
 G_3 ( v ) \equiv \sum_{I < J < K}^4 v_I v_J v_K \, , \\
 G_4 ( v ) & \equiv ( v_1 + v_3 ) ( v_2 + v_3 ) ( v_1 + v_4 ) ( v_2 + v_4 ) \, .
\eea
Then, the index can be written as
\bea
 \label{ABJM:TTI:unrefined}
 \frac{\log Z}{N^{3/2}} & =
 k \int \rd t \? \rho (t ) \left( t \delta \fp (t) + \delta v ( t) \fn (t) \right)
 + \mu \left( \int \rd t \? \rho ( t) - 1 \right)
 + \frac12 G_1( \fs ) \int \rd t \? \left( \rho (t ) \?  \delta v ( t) \right)^2 \\
 & - \int \rd t \? \rho (t )^2 \delta v ( t ) \left( G_1 ( \Delta ) \delta \fp ( t) + \sum_{I = 1}^4 \fs_I \frac{\partial G_2 ( \Delta )}{\partial \Delta_I} \right)
 + G_2 ( \Delta ) \int \rd t \? \rho (t )^2 \delta \fp ( t ) \\
 & - \frac12 \int \rd t \? \rho ( t)^2 \bigg[
 \sum_{\substack{I < J \\ (I , J \neq K)}}^4 \Delta_I \Delta_J \fs_K
+  \fn ' ( t ) \left( G_3 ( \Delta ) - G_1 ( \Delta ) \delta v ( t )^2 - G_2 ( \Delta ) \delta v ( t ) \right)
 \bigg]
 \, ,
\eea
where we included the Lagrange multiplier $\mu$ to enforce the normalization of $\rho ( t)$.

We define the inner segment as
\be
t_< ~ \text{ s.t. } ~ \delta v ( t_< ) = - \Delta_3 \;,\qquad\qquad t_> ~ \text{ s.t. } ~ \delta v ( t_> ) = \Delta_1  \, .
\ee
Schematically, we have:
\begin{center}
\begin{tikzpicture}[scale=2]
\draw (-2.,0) -- (2.,0);
\draw (-2,-.05) -- (-2, .05); \draw (-0.7,-.05) -- (-0.7, .05); \draw (0.7,-.05) -- (0.7, .05); \draw (2,-.05) -- (2, .05);
\node [below] at (-2,0) {$t_\ll$}; \node [below] at (-2,-.3) {$\rho=0$};
\node [below] at (-0.7,0) {$t_<$}; \node [below] at (-0.7,-.3) {$\delta v = - \Delta_3$}; \node [below] at (-0.7,-.6) {$\delta \fp = \fs_3$};
\node [below] at (0.7,0) {$t_>$}; \node [below] at (0.7,-.3) {$\delta v = \Delta_1$}; \node [below] at (0.7,-.6) {$\delta \fp = - \fs_1$};
\node [below] at (2,0) {$t_\gg$}; \node [below] at (2,-.3) {$\rho=0$};
\end{tikzpicture}
\end{center}
In the left segment, where $\delta v ( t ) = - \Delta_3$ and $\delta \fp ( t ) = \fs_3$,
the equations obtained from varying the index \eqref{ABJM:Z:factor} with respect to $\delta v ( t )$ and $\delta \fp ( t )$
need not hold,%
\footnote{This is a large $N$ effect. As explained in \cite{Benini:2015eyy}, they hold when including subleading exponential corrections.}
and we obtain
\be
 \begin{aligned}
 \rho ( t ) & = r_2 - \frac{k \Delta _3}{(\Delta _1+\Delta _3 ) (\Delta _2+\Delta _3 ) (\Delta _3-\Delta _4 )} \? t \, , \\
 \fn ( t ) & = \frac{\mu }{k \Delta _3} + \frac{\fs_3}{\Delta _3} \? t + \frac{c_2}{\rho (t)} 
 + \frac{\rho (t)}{2 k \Delta_3} \\
 & \times \bigg[
 \fs_1 (\Delta _3-\Delta _4 ) (\Delta _2+\Delta _3 )
 + \fs_2 (\Delta _1+\Delta _3 ) (\Delta _3-\Delta _4 )
 - \fs_4 (\Delta _1+\Delta _3 ) (\Delta _2+\Delta _3 ) \\
 & + \fs_3 \left( (\Delta _3-\Delta _4 )^2 + (3 \Delta _3-2 \Delta _4 ) G_1(\Delta ) + \left( 2 - \frac{\Delta _4}{\Delta _3}\right) G_2(\Delta )\right)
 \bigg] \, ,
 \end{aligned}
\ee
where $r_2$, $c_2$ are constants of integrations.
In the middle segment, we have
\be
 \begin{aligned}
 \rho ( t ) & = r_1 - k \frac{G_2( \Delta )}{G_4(\Delta )} \? t \, , \\
 \delta v ( t) & = \frac{1}{G_1 ( \Delta )} \left( \frac{k t}{\rho (t)} + G_2 ( \Delta ) \right) , \\
 \delta p ( t ) & = \frac{k}{G_1(\Delta ) \rho (t)} \left(t \left(\frac{G_1(\fs)}{G_1(\Delta )} + \fn'(t) \right) + \fn(t) \right) \\
 & + \frac{(\Delta _1+\Delta _3) (\Delta _2+\Delta _3) \fs_4 + (\Delta _1+\Delta _4 ) (\Delta _2+\Delta _4 ) \fs_3}{G_1( \Delta )^2} \\
 & - \frac{(\Delta _2+\Delta _4 ) (\Delta _2+\Delta _3 ) \fs_1 + (\Delta _1+\Delta _3 ) (\Delta _1+\Delta _4 ) \fs_2}{G_1( \Delta )^2} \, , \\
 \fn ( t ) & = \frac{c_1}{\rho ( t )} - \frac{ \mu}{k} \frac{G_1(\Delta )}{G_2 (\Delta )} - \left( \frac{\Delta _3 \fs_4 + \Delta_4 \fs_3 - \Delta _1 \fs_2 - \Delta _2 \fs_1}{G_2(\Delta )} + \frac{G_1(\fs)}{G_1(\Delta )} \right) t
 + \frac{\rho(t)}{2 k \? G_1( \Delta ) G_2( \Delta )^2} \\
 & \times \Big\{ \fs_1 ( \Delta _2+\Delta _3 ) (\Delta _2+\Delta _4 )\left[ G_1(\Delta ) \left(\left(G_1(\Delta )-\Delta _2\right) G_2(\Delta )-G_3(\Delta )\right)+2 G_4(\Delta ) \right]  \\
 & + \fs_2 (\Delta _1+\Delta _3 ) (\Delta _1+\Delta _4 ) \left[ G_1(\Delta ) \left(\left(G_1(\Delta )-\Delta _1\right) G_2(\Delta )-G_3(\Delta )\right)+2 G_4(\Delta ) \right] \\
 & + \fs_3 (\Delta _1+\Delta _4 ) (\Delta _2+\Delta _4 ) \left[ G_1(\Delta ) \left(\left(G_1(\Delta )-\Delta _4\right) G_2(\Delta )+G_3(\Delta )\right)-2 G_4(\Delta ) \right] \\
 & + \fs_4 (\Delta _1+\Delta _3 ) (\Delta _2+\Delta _3 ) \left[ G_1(\Delta ) \left(\left(G_1(\Delta )-\Delta _3\right) G_2(\Delta )+G_3(\Delta )\right)-2 G_4(\Delta ) \right] \Big\} \, ,
 \end{aligned}
\ee
where $r_1$, $c_1$ are constants of integrations.
In the right segment, extremizing the index with respect to $\rho ( t)$ and $\fn ( t )$
at fixed $( \delta v ( t ) ,  \delta p ( t ) ) = ( \Delta_1, - \fs_1)$, we find
\be
 \begin{aligned}
 \rho ( t ) & = r_3 + \frac{k \Delta _1}{(\Delta _1-\Delta _2 ) (\Delta _1+\Delta _3 ) (\Delta _1+\Delta _4 )} \? t \, , \\
 \fn ( t ) & = -\frac{\mu }{k \Delta _1}+\frac{\fs_1}{\Delta _1} \? t+ \frac{c_3}{\rho (t)}
 + \frac{\rho (t)}{2 k \Delta _1} \\
 & \times \bigg[
 \fs_2 (\Delta _1+\Delta _4 ) (\Delta _1+\Delta _3 ) - \fs_3 (\Delta _1 - \Delta _2 ) (\Delta _1+\Delta _4 ) - \fs_4 (\Delta _1 - \Delta _2 ) (\Delta _1 + \Delta _3 ) \\
 & - \fs_1 \left( (\Delta _1-\Delta _2 )^2 + ( 3 \Delta _1-2 \Delta _2 ) G_1(\Delta ) - \left( 2 - \frac{\Delta _2}{\Delta _1} \right) G_2(\Delta ) \right)
 \bigg] \, ,
 \end{aligned}
\ee
where $r_3$, $c_3$ are constants of integrations.
The support of $\rho(t)$ is then given by
\bea
 \rho ( t_{\ll} ) \big|_{\text{L.\?I.}} & = 0 \hspace{0.8cm} \Rightarrow
 && t_{\ll} = \frac{( \Delta_1 + \Delta_3 ) (\Delta_2 + \Delta_3 ) (\Delta_3 - \Delta_4 )}{k \Delta_3}  \? r_2 \, , \\
 \delta v ( t_< ) \big|_{\text{M.\?I.}} & = - \Delta_3 \hspace{0.8cm} \Rightarrow
 && t_{<} = - \frac{(\Delta _1+\Delta _3 ) (\Delta _2+\Delta _3 ) (\Delta _1+\Delta _4 ) (\Delta _2+\Delta _4 )}{k \Delta _4 G_1 (\Delta _4)} \? r_1 \, , \\
 \delta v ( t_> ) \big|_{\text{M.\?I.}} & = \Delta_1 \hspace{1.1cm} \Rightarrow
 && t_{>} = \frac{(\Delta _1+\Delta _3 ) (\Delta _2+\Delta _3 ) (\Delta _1+\Delta _4 ) (\Delta _2+\Delta _4 )}{k \Delta _2 G_1 (\Delta)} \? r_1 \, , \\
 \rho ( t_{\gg} ) \big|_{\text{R.\?I.}} & = 0 \ \hspace{1.1cm} \Rightarrow
 && t_{\gg} = - \frac{( \Delta_1 - \Delta_2 ) (\Delta_1 + \Delta_3 ) (\Delta_1 + \Delta_4 )}{k \Delta_1} \? r_3 \, .
\eea
Now, it has remained to determine the Lagrange multiplier $\mu$ and the constants of integrations $c_i$ and $r_i$, $i =1, 2, 3$.
First, note that the regularity of the magnetic flux $\fn ( t)$ at the endpoints of the support of $\rho ( t)$ yields
\bea
 \text{Coeff } [\fn ( t_{\ll} ) ,\rho(t), -1] = 0 \qquad \Rightarrow \qquad c_2 = 0 \, , \\
 \text{Coeff } [\fn ( t_{\gg} ) ,\rho(t), -1] = 0 \qquad \Rightarrow \qquad c_3 = 0 \, .
\eea
Second, the density $\rho ( t)$ and the magnetic flux $\fn ( t)$ are continuous piece-wise functions.
Thus, continuity of $\rho(t)$ fixes the constants $r_{23}$ and $r_3$,
\bea
 \rho ( t_< ) \big|_{\text{M.\?I.}} & = \rho ( t_< ) \big|_{\text{L.\?I.}}  \qquad \Rightarrow \qquad
 r_2 = -\frac{( \Delta _1+\Delta _4 ) ( \Delta _2+\Delta _4 )}{(\Delta _3-\Delta _4 ) G_1(\Delta )} \? r_1 \, , \\
 \rho ( t_> ) \big|_{\text{M.\?I.}} & = \rho ( t_> ) \big|_{\text{R.\?I.}} \qquad \Rightarrow \qquad
 r_3 = - \frac{(\Delta _2+\Delta _3 ) (\Delta _2+\Delta _4 )}{(\Delta _1-\Delta _2 ) G_1 ( \Delta )} \? r_1 \, .
\eea
Observe that
\be
  \delta v ( t_\ll ) \big|_{\text{M.\?I.}} = - \Delta_4 \, , \qquad \delta v ( t_\gg ) \big|_{\text{M.\?I.}} = \Delta_2 \, .
\ee
The continuity of $\fn(t)$ gives
\bea
 \mu & = \frac{( \Delta _1+\Delta _3 ) (\Delta _2+\Delta _3 ) (\Delta _1+\Delta _4 ) (\Delta _2+\Delta _4 )}{2 G_1(\Delta )} \? r_1 \sum_{I = 1}^4 \frac{\fs_I}{\Delta_I} \, , \\
 c_1 & = \left(\frac{(\Delta _1+\Delta _3 ) (\Delta _2+\Delta _3 ) (\Delta _1+\Delta _4 ) (\Delta _2+\Delta _4 )}{ \sqrt{2 k} \? (\Delta _1 \Delta _2-\Delta _3 \Delta _4 ) G_1( \Delta )} \? r_1 \right)^2
 \bigg( \sum_{I = 3}^4 \frac{\fs_I}{\Delta_I} - \sum_{I = 1}^2 \frac{\fs_I}{\Delta_I} \bigg) \, .
\eea
Finally, the normalization of $\rho ( t)$ fixes $r_1$,
\be
 \int \rd t \? \rho ( t ) = 1 \qquad \Rightarrow \qquad r_1 = \frac{G_1 (\Delta) \sqrt{2 k \Delta _1 \Delta _2 \Delta _3 \Delta _4}}{(\Delta _1+\Delta _3 ) (\Delta _2+\Delta _3 ) (\Delta _1+\Delta _4 ) (\Delta _2+\Delta _4 )} \, .
\ee
Plugging back the saddle point into the unrefined index \eqref{ABJM:TTI:unrefined}, we obtain
\be
 \log Z = - \frac23 N^{3/2} \mu = - \frac{N^{3/2}}{3} \sqrt{2 k \Delta _1 \Delta _2 \Delta _3 \Delta _4} \sum_{I = 1}^4 \frac{\fs_I}{\Delta_I} \, ,
\ee
in precise agreement with \cite[(2.89)]{Benini:2015eyy}.

\section{Derivation of general rules for theories with $N^{5/3}$ scaling of the index}
\label{app:N5/3}

We consider the following ansatz for the large $N$ saddle point eigenvalue distribution
\be
 \label{app:N^5/3:ansatz}
 u^{(a)} (t) = N^{1/3} ( \ii t + v(t) ) \, , \qquad \fm ( t ) = \ii N^{1/3} \fn(t) \, .
\ee
Furthermore, we assume that
\be
 k_{\text{CS}} \equiv \sum_{a = 1}^{|\cG|} k_a \neq 0 \, ,
\ee
that corresponds to turning on the Romans mass $F_0$ in the dual type IIA supergravity \cite{Gaiotto:2009mv}.
\subsection{Chern-Simons}

Using the scaling ansatz \eqref{app:N^5/3:ansatz}, each gauge group $a$ with CS level $k_a$ contributes to the twisted index as
\be
 \label{CS:largeN}
 \log Z_{\text{CS}} = \ii k_a \sum_{i = 1}^{N} \fm_i u_i  \overset{N \gg 1}{=} - k_a N^{5/3} \int \rd t \? \rho ( t ) \fn ( t ) ( \ii t + v ( t ) ) \, .
\ee

\subsection{Chiral multiplet}

Let us evaluate the large $N$ contribution of a bi-fundamental chiral multiplet
transforming in a representation $(\overline{\bf N} , {\bf N})$ of $\U(N)_a \times \U(N)_b$
and with chemical potential and magnetic flux $(\Delta, \fs)$ to the refined twisted index.%
\footnote{The contribution of a chiral multiplet in the adjoint representation of $\U(N)_a$ can be derived in exactly the same fashion.}
As before, we break the sum $\sum_{i , j = 1}^N $ in \eqref{logZ:chiral:dilog} into $\sum_{i < j} + \sum_{i > j} + ( i \to j )$.
The contribution coming from $( i \to j)$ is subleading.
Let us evaluate \eqref{logZ:chiral:dilog:expand:i<j} in the large $N$ limit.
We find
\be
 \label{Z:chi:i<j}
 \log Z_{(b , a)}^{i < j} = N^2 \sum_{n = 1}^{\infty} \frac{e^{\ii n ( \Delta + \frac{\epsilon}{2} \fs )}}{n \left( e^{\ii n \epsilon} - 1\right)} \int \rd t \? \rho ( t ) I_n ( t ) \, ,
\ee
where we defined
\bea
 I_n ( t ) & \equiv \int_t \rd t' \? \rho ( t' ) e^{- n N^{1/3} ( t' - t)}
 \left( e^{\ii n \epsilon \left( \ii N^{1/3} \left( \fn ( t' ) - \fn ( t ) \right) - \fs + 1 \right)} - 1 \right) 
 e^{\ii n N^{1/3} \left( v( t' ) - v ( t ) - \frac{\ii}{2} \epsilon \left( \fn ( t' ) - \fn ( t ) \right) \right)} \\
 & \equiv \int_t \rd t' \? \cI_n ( t', t ) \, .
\eea
Performing integration by parts we obtain
\be
 \label{int:by:parts:1}
 \begin{aligned}
  I_n ( t ) & \overset{N\gg1}{=} - \frac{N^{-1/3}}{n} \rho ( t' )  \cI_n ( t', t ) \Big|_t \\
  & + \int_t \rd t' \? \rho ( t' ) \cI_n ( t', t )
  \left[ \ii v'( t' ) + \frac{\ii \epsilon}{2} \fn'( t' ) \cot \left( \frac{n \epsilon}{2} \left( \ii N^{1/3} ( \fn ( t' ) - \fn ( t) ) - \fs + 1 \right) \right)
  \right] .
 \end{aligned}
\ee
We again perform integration by parts on the second line of \eqref{int:by:parts:1} and find, at large $N$,
\bea
  I_n & = - \frac{N^{-1/3}}{n} \rho ( t' ) \cI_n ( t', t ) 
  \left[ 1 + \ii v'( t' ) + \frac{\ii \epsilon}{2}  \fn' ( t' ) \cot \left( \frac{n \epsilon}{2} \left( \ii N^{1/3} ( \fn ( t' ) - \fn ( t ) ) - \fs + 1 \right) \right) \right]_t \\
  & - \int_t \rd t' \? \rho ( t' ) \cI_n ( t', t )
  \left[ v'( t' )^2 - \frac{\epsilon^2}{4} \fn' ( t' )^2 + \epsilon \? \fn'( t' ) v'( t' ) \cot \left( \frac{n \epsilon}{2} \left( \ii N^{1/3} ( \fn ( t' ) - \fn ( t ) ) - \fs + 1 \right) \right) \right] .
\eea
Via a repeated application of integration by parts we can then write
\bea
 I_n ( t ) & = \frac{N^{-1/3}}{n} \rho( t )
 \left( e^{\ii n \epsilon ( 1 - \fs )} - 1 \right)
 \sum_{l, r = 1}^\infty \frac{(1 + \ii ) \left( 1 + \ii (-1)^r  \right) ( l - r + 2)_{r-1}}{2^r ( r - 1)!}
 \left( \ii v'(t) \right)^{l - r +1} \\
 & \hspace{5.5cm} \times \left( \epsilon \fn'(t) \right)^{r-1}
 \cot^{\frac{1}{2} \left(1 + (-1)^r \right)} \left( \frac{n \epsilon}{2} ( 1 - \fs ) \right) ,
\eea
where we only kept the leading order terms that contribute to the large $N$ twisted index.
Here, $(x)_n$ is the Pochhammer symbol.
Observe that the sum over $l$ can be done explicitly and thus we obtain
\bea
 I_n & = \frac{N^{-1/3}}{n} \rho ( t )
 \left( e^{\ii n \epsilon ( 1 - \fs )} - 1 \right)
 \sum_{r = 1}^\infty \frac{\ii^{r-1} (-1)^{\left \lfloor \frac{r-1}{2} \right\rfloor}}{2^{r-1}}
 \frac{\left( \epsilon \fn'(t) \right)^{r-1}}{\left( 1 - \ii v'(t) \right)^{r}}
 \cot^{\frac{1}{2} \left(1 + (-1)^r \right)} \left( \frac{n \epsilon}{2} ( 1 - \fs ) \right) .
\eea
The sum over $r$ can also be done and we find the following compact expression for $I_n$
\bea
 I_n ( t ) & = \frac{N^{-1/3}}{n} \rho( t )
 e^{\frac{\ii n \epsilon}{2} ( 1 - \fs )}
 \left(
 \frac{e^{\frac{\ii n \epsilon}{2} ( 1 - \fs )}}{\left(1-i v'(t)\right) + \frac{\epsilon }{2} \fn'(t)}
 - \frac{e^{- \frac{\ii n \epsilon}{2} ( 1 - \fs )}}{\left(1-i v'(t)\right) - \frac{\epsilon }{2} \fn'(t)}
 \right) .
\eea
Therefore, in the large $N$ limit, \eqref{Z:chi:i<j} is simplified to
\be
 \label{ini:chiral:final:i<j}
 \log Z_{(b , a)}^{i<j} = N^{5/3} \sum_{n = 1}^{\infty}
 \frac{e^{\frac{\ii n \epsilon}{2}}}{n^2 \left( e^{\ii n \epsilon} - 1 \right)}
  \int \rd t \? \rho( t )^2
 \left(
 \frac{e^{\ii n \left( \Delta + \frac{\epsilon}{2} ( 1 - \fs  ) \right)}}{\left(1-i v'(t)\right) + \frac{\epsilon }{2} \fn'(t)}
 - \frac{e^{\ii n \left( \Delta - \frac{\epsilon}{2} ( 1 - \fs ) \right)}}{\left(1-i v'(t)\right) - \frac{\epsilon }{2} \fn'(t)}
 \right) .
\ee
Now, it remains to perform the sum over $n$. We obtain
\be
 \label{chiral:final:i<j}
 \log Z_{(b , a)}^{i<j} = - N^{5/3}
 \sum_{n = 1}^\infty
  \int \rd t \? \rho( t )^2
 \left(
 \frac{\Li_2 \left( e^{\ii \left( \Delta + \epsilon (n - \frac{\fs}{2}) \right)} \right)}{\left(1-i v'(t)\right) + \frac{\epsilon }{2} \fn'(t)}
 - \frac{\Li_2 \left( e^{\ii \left( \Delta + \epsilon (n - 1 + \frac{\fs}{2}) \right)} \right)}{\left(1-i v'(t)\right) - \frac{\epsilon }{2} \fn'(t)}
 \right) .
\ee
The summation $\sum_{i > j}$ is similar to \eqref{chiral:final:i<j} and it reads
\be
 \label{chiral:final:i>j}
 \log Z_{(b , a)}^{i>j} = N^{5/3}
 \sum_{n = 1}^\infty
  \int \rd t \? \rho( t )^2
 \left(
 \frac{\Li_2 \left( e^{- \ii \left( \Delta - \epsilon (n - 1 + \frac{\fs}{2}) \right)} \right)}{\left(1-i v'(t)\right) + \frac{\epsilon }{2} \fn'(t)}
 - \frac{\Li_2 \left( e^{- \ii \left( \Delta - \epsilon (n - \frac{\fs}{2}) \right)} \right)}{\left(1-i v'(t)\right) - \frac{\epsilon }{2} \fn'(t)}
 \right) .
\ee
Combining \eqref{chiral:final:i<j} and \eqref{chiral:final:i>j}, we finally arrive at the following compact expression for the
contribution of a bi-fundamental chiral multiplet to the refined twisted index at large $N$
\be
 \label{chiral:FINAL}
 \log Z_{(b , a)} = N^{5/3}
  \int \rd t \? \rho( t )^2
 \left(
 \frac{\psi \left( - \Delta + \epsilon \? \frac{\fs}{2} ; \epsilon \right)}{\left(1-i v'(t)\right) + \frac{\epsilon }{2} \fn'(t)}
 - \frac{\psi \left(  -\Delta + \epsilon \left( 1 - \frac{\fs}{2} \right) ; \epsilon \right)}{\left(1-i v'(t)\right) - \frac{\epsilon }{2} \fn'(t)}
 \right) .
\ee

\paragraph*{$\epsilon = 0$ case --}
Taking the $\epsilon \to 0$ limit of \eqref{ini:chiral:final:i<j} we easily obtain
\bea
 \label{noepsilon:i<j}
 \log Z_{(b , a)}^{i<j} & = N^{5/3} \sum_{n = 1}^{\infty}
 \frac{e^{\ii n \Delta}}{n^3}
 \int \rd t \? \rho( t )^2
 \left( n \frac{1 - \fs}{ 1 - \ii v'(t)} + \ii \frac{\fn'(t)}{\left( 1 - \ii v'(t) \right)^2} \right) \\
 & = N^{5/3} \int \rd t \? \rho( t )^2
 \left(  \Li_2 \left( e^{\ii \Delta} \right) \frac{ 1- \fs}{ 1 - \ii v'(t)} + \ii  \Li_3 \left( e^{\ii \Delta} \right) \frac{\fn'(t)}{\left( 1 - \ii v'(t) \right)^2} \right) .
\eea
The summation $\sum_{i > j}$ is similar to \eqref{noepsilon:i<j} and it reads
\be
 \label{noepsilon:i>j}
 \log Z_{(b , a)}^{i > j} = N^{5/3} \int \rd t \? \rho( t )^2
 \left(  \Li_2 \left( e^{- \ii \Delta} \right) \frac{1 - \fs}{ 1 - \ii v'(t)} - \ii  \Li_3 \left( e^{- \ii \Delta} \right) \frac{\fn'(t)}{\left( 1 - \ii v'(t) \right)^2} \right) .
\ee
Combining \eqref{noepsilon:i<j} and \eqref{noepsilon:i>j}, and using the inversion formul\ae\;\eqref{Li:inversion:g}
for $0 < \re ( \Delta ) < 2 \pi$, we finally obtain the contribution of a chiral multiplet to the large $N$ twisted index
\be
 \label{noepsilon:chiral:largeN}
 \log Z_{(b , a)} \big|_{\epsilon = 0} = N^{5/3} \int \rd t \? \rho( t )^2
 \left(  g_2 ( \Delta ) \frac{1 - \fs}{ 1 - \ii v'(t)} - g_3 ( \Delta ) \frac{\fn'(t)}{\left( 1 - \ii v'(t) \right)^2} \right) .
\ee
\paragraph*{Asymptotic expansion around $\epsilon = 0$ --}
Starting with \eqref{chiral:FINAL} we can write down the following asymptotic expansion around $\epsilon = 0$
\bea
 \label{chiral:asymp:ini}
 \log Z_\chi = - N^{5/3} \int \rd t \? \rho ( t )^2
 \Bigg\{
 & \bigg[
 g_3( \Delta ) \frac{\fn' ( t )}{\left( 1 - \ii v'( t ) \right)^2}
 + \frac{g_2 ( \Delta ) g_1 ( \pi \fs )}{\pi} \frac{1}{1 - \ii v' ( t )}
 + \frac{g_1( \Delta ) g_2 ( \pi \fs )}{\pi^2} \fn' ( t )^{-1} \\
 & + \frac{g_3 ( \pi \fs )}{\pi^3} \frac{\fn' ( t )^{-2}}{\left( 1 - \ii v'(t) \right)^{-1}} \bigg]
  \sum_{n = 0}^{\infty} \left(\frac{\epsilon \? \fn' ( t )}{2 \left( 1 - \ii v' ( t ) \right)} \right)^{2 n} \\
 & - \frac{g_3 ( \pi \fs )}{\pi^3} \frac{\fn' ( t )^{-2}}{\left( 1 - \ii v' ( t ) \right)^{-1}}
 - \frac{g_1 ( \Delta ) g_2 ( \pi \fs )}{\pi^2} \fn' ( t )^{-1} \Bigg\} \, .
 \eea
Using $\sum_{n = 0}^\infty x^{2 n} = (1 - x^2)^{-1}$, \eqref{chiral:asymp:ini} is then simplified to
\be
\label{cardy:FINAL:chi}
\log Z_{(b , a)} = N^{5/3} \int \rd t \? \rho ( t )^2
 \left(
 \frac{\cG ( \Delta, \fs , \epsilon )}{\left (1 - \ii v' ( t ) \right) + \frac{\epsilon}{2} \fn' ( t )}
 - \frac{\cG ( \Delta, 2 - \fs, \epsilon )}{\left(1 - \ii v' ( t ) \right) - \frac {\epsilon} {2} \fn' ( t )}
 \right) ,
\ee
where $\cG ( \Delta, \fs , \epsilon )$ is given in \eqref{def:cG:poly}.
Let us define the \emph{equivariant chemical potentials}
\bea
 & w^{(1)} ( t ) \equiv \left( \ii t + v ( t ) \right) + \frac{\ii \epsilon}{2} \fn ( t ) \, , && \bbDelta_I^{(1)} \equiv \frac{1}{\omega} \left( \Delta_I + \pi ( \omega - 1 ) + \frac{\epsilon}{2} ( 1 - \fs_I ) \right) , \\
 & w^{(2)} ( t ) \equiv \left( \ii t + v ( t ) \right) - \frac{\ii \epsilon}{2} \fn ( t ) \, , \quad && \bbDelta_I^{(2)} \equiv \frac{1}{\omega} \left( \Delta_I + \pi ( \omega - 1 ) - \frac{\epsilon}{2} ( 1 - \fs_I ) \right) ,
\eea
with $\omega$ as before, see \eqref{def:omega:epsilon}.
Then, \eqref{cardy:FINAL:chi} takes the following remarkable factorized form
\be
\label{cardy:factorized:chi}
\log Z_{(b , a)} = \frac{\ii \omega^3}{\epsilon} N^{5/3} \int \rd t \? \rho ( t )^2
 \left(
 \frac{g_3 ( \bbDelta^{(1)} )}{w'^{(1)}( t )}
 - \frac{g_3 ( \bbDelta^{(2)} )}{w'^{(2)}( t )}
 \right) ,
\ee
reproducing \eqref{main:cardy:factorized:chi} after setting $\bbDelta \equiv \bbDelta_{(a,b)}$ and $\fs \equiv \fs_{(a,b)}$.

\subsection{Vector multiplet}

The contribution of a vector multiplet can be simply obtained by using \eqref{chi:map:vec}.
In particular, from \eqref{cardy:FINAL:chi} we obtain the following expression for a vector multiplet
\be
\label{cardy:FINAL:vec}
\log Z_{\cV} = \frac{\pi}{12} N^{5/3} \int \rd t \? \rho ( t )^2
 \left(
 \frac{\epsilon - 2 \pi}{\left (1 - \ii v' ( t ) \right) + \frac{\epsilon}{2} \fn' ( t )}
 - \frac{\epsilon + 2 \pi}{\left(1 - \ii v' ( t ) \right) - \frac {\epsilon} {2} \fn' ( t )}
 \right) ,
\ee
that is \eqref{main:N^5/3:cardy:FINAL:vec}.

\section{Large $N$ factorization and holomorphic blocks}
\label{app:gangnam} 

In this section we sketch the derivation of formula \eqref{Bethehol2} for generic theories with $N^{3/2}$ scaling, generalizing \cite{Choi:2019dfu}.
A similar analysis could be done for theories with $N^{5/3}$ scaling.

Consider the contribution \eqref{logZ:chiral:ini} to the index of a chiral bi-fundamental multiplet. It can be obtained by gluing two holomorphic blocks \cite{Pasquetti:2011fj,Beem:2012mb,Nieri:2015yia}
\bea
 \label{holB}
 Z_{(b,a)}=\prod_{\sigma=1}^2 B \left( u^{(\sigma)} ; \bar \Delta^{(\sigma)} | \epsilon^{(\sigma)} \right) \, ,
\eea
where%
\footnote{Here $(x ; q)_\infty =\prod_{n=0}^\infty (1 - x q^n)$ for $|q|<1$  is the $q$-Pochhammer symbol.
The formula $(x ; q^{-1})_\infty =1/(q x ; q)$ is used to define the $q$-Pochhammer symbol for $|q|>1$.}
\bea
 B(u ; \Delta | \epsilon ) =  e^{\frac{\ii}{2\epsilon} g_2(u+\Delta) -\frac{\ii(r-1)}{4} g_1(u+\Delta)}  (q^{1-r/2}\? x\? y ; q)_\infty \, ,
\eea
with $x=e^{\ii u}$, $y=e^{\ii \Delta}$, $q=e^{\ii \epsilon}$, and $r$ is the R-symmetry charge.
The gluing rules are
\bea
 \label{A-twisted:var2}
 u^{(\sigma)}_i & \equiv u_i + \frac{\epsilon^{(\sigma)}}{2} \fm_i \, , \qquad \qquad \bar \Delta^{(\sigma)}_I = \Delta_I + \frac{\epsilon^{(\sigma)}}{2} \ft_I \, ,\qquad \qquad  \sigma = 1,2,\\
 \epsilon^{(1)} & \equiv \epsilon \, , \hspace{3.7cm} \epsilon^{(2)} \equiv - \epsilon \, ,
\eea
where $\ft$ are flavor fluxes satisfying $\sum_{I\in W_a} \ft_I = 0$, with $W_a$ denoteing a generic monomial term in the superpotential.
They are related to the set of fluxes $\fs_I$ used in the main text, and satisfying $\sum_{I\in W_a} \fs_I = 2$, via $\fs_I = r_I - \ft_I$.

Stripping off the exponential terms in \eqref{holB}, which only enter in the long-range forces cancellation as before, the two blocks can be written
in our favorite parameterization \eqref{A-twisted:var} as
\bea 
 B \left( u^{(\sigma)} ; \bar \Delta^{(\sigma)} | \epsilon^{(\sigma)} \right) = \left( q^{(\sigma)}\? x^{(\sigma)}\? y^{(\sigma)} ; q^{(\sigma)} \right)_\infty\equiv \left(x^{(\sigma)}\? y^{(\sigma)} ; 1/q^{(\sigma)} \right)_\infty^{-1} \, .
\eea

In the factorization method we introduce two set of quantities $\rho^{(\sigma)} ( t )$ and $\delta v^{(\sigma)}( t )$ for the independent variables $u^{(\sigma)}_i$.
Focusing on $\sigma=1$ and restricting the analysis to an asymptotic expansion in $\epsilon$%
\footnote{We use the formula $\log ( x ; q^{-1})_\infty =\sum_{s=0}^\infty (-\ii \epsilon)^{s-1} \frac{B_s}{s!} \Li_{2-s}( x )$ valid for $|q^{-1}|<1$. In this appendix we assume that $\epsilon$ has a small negative imaginary part. To use the same expansion for the case $\sigma=2$, we need first to use the inversion formula given in the previous footnote.} 
\bea
 \log B^{(1)} = - \sum_{s=0}^\infty (-\ii \epsilon)^{s-1} \frac{B_s}{s!} \Li_{2-s}(x^{(1)}\, y^{(1)}) \, ,
\eea
where $B_s = \left\{1, - \frac{1}{2} , \frac{1}{6} , 0 , - \frac{1}{30} , 0 , \ldots \right\}$ is the $s$th Bernoulli number.
At large $N$, each term in this expansion can be treated as in \cite{Benini:2015eyy,Hosseini:2016tor}.
The steps are similar to the previous computations and we refer to \cite{Benini:2015eyy,Hosseini:2016tor} for details.
We split the gauge sum into  $\sum_{i < j}$ and  $\sum_{i > j}$, use the inversion formul\ae\;\eqref{Li:inversion:g} in the second term and perform the large $N$ limit.
The final result, up to polynomial terms that enter in cancelling the long-range forces, is
\bea
 \label{interW}
 \frac{\log B^{(1)}}{N^{3/2}} &= - \sum_{s=0}^\infty (-\ii \epsilon)^{s-1} \frac{B_s}{s!} \int \rd t \? \rho^{(1)}( t )^2 \Big( \Li_{3-s}(e^{\ii(\delta v^{(1)} (t ) +\Delta^{(1)})})
 + (-1)^{3-s} \Li_{3-s}(e^{-\ii(\delta v^{(1)} (t ) +\Delta^{(1)})}) \Big) \\
 & = - \sum_{s=0}^\infty (- \epsilon)^{s-1} \frac{B_s}{s!} \int \rd t \rho^{(1)}( t )^2 g_{3-s} (\delta v^{(1)} (t ) +\Delta^{(1)}) \, ,
\eea
where we used again \eqref{Li:inversion:g}. 
Since $g_{s<0} (u) \equiv 0$, only the first few terms in this sum contribute.
The leading term for $\epsilon \to 0$ is  the contribution of a bi-fundamental chiral field to the twisted  superpotential \cite{Benini:2015eyy,Hosseini:2016tor}%
\footnote{See \cite[(3.17)]{Hosseini:2016tor}, noting that $\wt \cW_{\text{here}} = - \cV_{\text{there}}$, or compare, for example, \eqref{ABJM:effective:W2} for the ABJM theory.} 
\bea
 \label{Whom}
 \frac{\ii}{\epsilon N^{3/2}} \wt \cW (\rho ( t ), \delta v (t ) ; \Delta) \bigg|_{\text{chiral}} = \frac{1}{ \epsilon} \int \rd t \? \rho( t )^2 g_{3} (\delta v (t ) +\Delta) \, ,
\eea
in agreement with   \eqref{Bethehol}.

The contribution of a vector multiplet is  obtained from \eqref{interW} by setting $\delta v ( t )=0$ and $\Delta=2 \pi$ and $\fs=2$. 
The contribution of the classical terms and (anti)-fundamental fields can be easily computed.

Including all the contributions from all the fields (and the classical terms) and keeping all perturbative orders in $\epsilon$, we obtain the $\sigma=1$ contribution to \eqref{Bethehol2}
\bea
 \label{finalW}
 \log B^{(1)} \equiv \frac{\ii}{\epsilon^{(1)}} \wt \cW_{\text{hom}} (\rho^{(1)}( t ) , \delta v^{(1)} (t ) ; \Delta^{(1)})\, ,
\eea
where $\wt \cW_{\text{hom}}$ is the full effective twisted superpotential  computed in \cite{Benini:2015eyy,Hosseini:2016tor} written  as a homogeneous function of its variables. This can be done by using the constraints    $\sum_{I\in W_a} \Delta_I =2 \pi$ to replace all occurrences  of $\pi$ in  $\wt \cW$. The explicit dependence on $\epsilon$ emerges from the replacement of the original variables with the equivariant counterparts \eqref{A-twisted:var}.
The final result is correct only if using the homogeneous form of $\wt \cW$ since now
$\sum_{I\in W_a} \Delta_I ^{(1)}=2 \pi -\epsilon^{(1)}$.\footnote{Equivalently, we could use  the expression for $\wt \cW$ found in \cite{Benini:2015eyy,Hosseini:2016tor} by replacing the original variables with the equivariant counterpart \eqref{A-twisted:var} and all occurences
of $\pi$ with $\pi -\frac{\epsilon^{(1)}}{2}$.}

The final formula \eqref{finalW} is valid for the class of theories with $N^{3/2}$ scaling considered in this paper.
The difference\footnote{Notice that vectors do not contribute  to $\wt\cW$ \cite{Benini:2015eyy,Hosseini:2016tor}.}
\bea
 \sum_{\text{vectors}} \log B^{(1)} \bigg|_{\eqref{interW}} + \sum_{\text{bi-fund}} \log B^{(1)} \bigg|_{\eqref{interW}} - \frac{\ii}{\epsilon} \wt \cW_{\text{hom}} \bigg|_\text{bi-fund} \, ,
\eea
is indeed proportional to
\bea
 - \frac{\pi}{6} \sum_{\substack{I = (b , a): + \\ I = (a , b): -}} \left( \pm \delta v^{(1)} (t) \right )
 - \frac{\pi^2}{6} \Bigg[ \sum_{\text{I (bi-fund)}} \Big( \frac{\Delta_I }{\pi}-1 \Big) + |\cG| \Bigg]
 + \frac{\pi \epsilon}{12} \Bigg[\sum_{\text{I (bi-fund)}} ( \fs_I-1 ) + |\cG| \Bigg] \, .
\eea
The first term in this expression vanishes because for each bi-fundamental connecting two gauge groups we have an anti-bi-fundamental connecting the same groups.
The second and third term cancel because $\Delta_I/\pi$ and $\fs_I$ can be regarded as R-charge assignments for the fields and, at large $N$,
we have $\Tr R=0$ for all R-symmetries, as it follows from \eqref{long-range2}.
The analysis of the classical terms and (anti)-fundamental fields is easy and it is left to the reader.

The case $\sigma=2$ is completely analogous.%
\footnote{Taking into account the different definitions of the Pochhammer symbol for $|q|<1$ and $|q>1$.}

\bibliographystyle{ytphys}

\bibliography{NoBAES}

\end{document}